\documentclass[pra,reprint,superscriptaddress,onecolumn,showpacs,preprintnumbers,amsmath,amssymb]{revtex4-1}
\usepackage{graphicx} 
\usepackage{dcolumn}  
\usepackage{amssymb,latexsym}
\usepackage{amsbsy} 

\usepackage{amsmath}

\newcommand \trace[1] {\mathrm{Tr}({#1})}
\DeclareMathOperator{\co}{co}
\DeclareMathOperator{\dom}{dom}
\newcommand \subdiff {\underline{\partial}}
\newcommand \supdiff {\bar{\partial}}

\newcommand \op[1] {\hat{#1}}    
\newcommand \bra[1] {\langle {#1} |}
\newcommand \ket[1] {| {#1} \rangle}

\newcommand \gaugeeq {\sim}
\newcommand \notgaugeeq {\nsim}

\begin{document}

\preprint{APS/123-QED}

\title{The choice of basic variables in current-density functional theory}

\author{Erik I. Tellgren}
\email{erik.tellgren@kjemi.uio.no}
\affiliation{Centre for Theoretical and Computational Chemistry, Department of Chemistry, University of Oslo, P.O. Box 1033 Blindern, N-0315 Oslo, Norway}
\author{Simen Kvaal}
\affiliation{Centre for Theoretical and Computational Chemistry, Department of Chemistry, University of Oslo, P.O. Box 1033 Blindern, N-0315 Oslo, Norway}
\author{Espen Sagvolden}
\affiliation{Centre for Theoretical and Computational Chemistry, Department of Chemistry, University of Oslo, P.O. Box 1033 Blindern, N-0315 Oslo, Norway}
\author{Ulf Ekstr\"om}
\affiliation{Centre for Theoretical and Computational Chemistry, Department of Chemistry, University of Oslo, P.O. Box 1033 Blindern, N-0315 Oslo, Norway}
\author{Andrew M. Teale}
\affiliation{Centre for Theoretical and Computational Chemistry, Department of Chemistry, University of Oslo, P.O. Box 1033 Blindern, N-0315 Oslo, Norway}
\affiliation{School of Chemistry, University of Nottingham, University Park, Nottingham, NG7 2RD, United Kingdom}
\author{Trygve Helgaker}
\affiliation{Centre for Theoretical and Computational Chemistry, Department of Chemistry, University of Oslo, P.O. Box 1033 Blindern, N-0315 Oslo, Norway}

\date{\today}

\begin{abstract}
  The selection of basic variables in current-density functional
  theory and formal properties of the resulting formulations are
  critically examined. Focus is placed on the extent to which the
  Hohenberg--Kohn theorem, constrained-search approach and Lieb's
  formulation (in terms of convex and concave conjugation) of standard
  density-functional theory can be generalized to provide foundations
  for current-density functional theory. For the well-known case with
  the gauge-dependent paramagnetic current density as a basic
  variable, we find that the resulting total energy functional is not
  concave. It is shown that a simple redefinition of the scalar
  potential restores concavity and enables the application of convex 
  analysis and convex/concave conjugation. As a result, the solution
  sets arising in potential-optimization problems can be given a
  simple characterization. We also review attempts to establish
  theories with the physical current density as a basic
  variable. Despite the appealing physical motivation behind this
  choice of basic variables, we find that the mathematical foundations
  of the theories proposed to date are unsatisfactory. Moreover, the
  analogy to standard density-functional theory is substantially
  weaker as neither the constrained-search approach nor the convex
  analysis framework carry over to a theory making use of the physical
  current density.
\end{abstract}

\pacs{Valid PACS appear here}
\maketitle

\section{\label{secINTRO}Introduction}

Density-functional theory (DFT) constitutes one of the most popular
methods in quantum chemistry. The foundations of DFT rest in
particular on three contributions: First, the Hohenberg--Kohn (HK) theorems
established a one-to-one mapping between a set of scalar potentials
and a set of ground-state densities as well as a variation principle
based on the density~\cite{HOHENBERG_PR136_864}. Here, the density is
the charge density (strictly, the negative of the charge density in units of the
elementary electron charge). Second, the Levy--Lieb
constrained-search expression provided a formal but explicit
expression for the intrinsic energy (the universal density functional)
and clarified significant fundamental
points~\cite{LEVY_PNAS76_6062}. Third, Lieb further generalized the
universal functional to a convex functional represented in terms of a
Legendre--Fenchel transform~\cite{LIEB_IJQC24_243}. From a
mathematical point of view, Lieb's formulation is particularly
attractive as it allows application of convex analysis to establish
several properties of the intrinsic energy
functional~\cite{ESCHRIG03,KUTZELNIGG_JMSTHEO768_163,AYERS_PRA73_012513}. Additionally,
Lieb's framework has made feasible practical calculations of
approximations to the exact intrinsic energy functional and adiabatic
connection curves~\cite{Harris:1974p1782,LANGRETH:1975p1425,GUNNARSSON:1976p1781,Gunnarsson:1977,LANGRETH:1977p1780},
enabling detailed comparisons of the properties of approximate
and near-exact density
functionals to be made~\cite{COLONNA_JCP110_2828,WU_JCP118_2498,TEALE_JCP130_104111,TEALE_JCP132_164115,TEALE_JCP133_164112,STROMSHEIM_JCP135_194109}.

Standard DFT, involving universal energy functionals of only the
charge density, is limited to the treatment of physical systems that
may be represented as eigenstates of Hamiltonians that differ only in
their scalar potentials. To treat systems subject to an external
magnetic field, it is necessary to introduce an additional dependence
on the magnetic field or its associated vector potential into the
Hamiltonian. Consequently, a dependence on a corresponding variable
apart from charge density is needed in the universal energy functional.
In magnetic-field density functional theory (B-DFT) this is
resolved by constructing a family of density functionals---one for
each external magnetic
field\cite{GRAYCE_PRA50_3089,SALSBURY_JCP107_7350}.

In the present work, we consider the alternative current density-functional theory (CDFT), where the additional variable is
either the paramagnetic current density or the physical current
density. We restrict our attention to non-relativistic formulations and
most of the discussion will for simplicity not be concerned with
densities or density-contributions arising from spin-degrees of
freedom. We term the variables on which the energy
functionals explicitly depend the \textit{basic variables} and
make a distinction between \textit{basic densities} and \textit{basic
  potentials}.  Many choices of basic densities are
conceivable~\cite{HIGUCHI_PRB69_035113,AYERS_PRA80_032510};
we require only that the choices result in useful density-functional
theories. Our perspective thus differs from that in recent works on CDFT by Pan and
Sahni~\cite{PAN_IJQC110_2833,PAN_JPCS73_630,SAHNI_PRA85_052502,PAN_PRA86_042502}, who restrict
the term \emph{basic variable} to
variables that admit an HK theorem.  Although it appears naturally in the
generic framework introduced by Ayers and
Fuentealba~\cite{AYERS_PRA80_032510}, the possibility of choosing
basic potentials other than the standard electromagnetic potentials
and fields has not previously been explored in detail.

By far the most developed form of CDFT is that due to Vignale and
Rasolt~\cite{VIGNALE_PRL59_2360,VIGNALE_PRB37_10685}, who use the
charge and paramagnetic current densities as basic variables.  For
these variables, a Kohn--Sham approach has been
formulated~\cite{VIGNALE_PRL59_2360} with an associated adiabatic-connection~\cite{LIU_PRA54_1328}, virial and scaling
relations~\cite{ERHARD_PRA53_R5,LIU_PRA54_1328,HIGUCHI_PRB65_195122,HIGUCHI_PRB75_195114} analogous to standard
Kohn--Sham DFT.  In addition, optimized-effective-potential (OEP)
approaches based on this formulation of CDFT have been presented to
treat non-collinear
magnetism~\cite{LEE_PRA59_209,PITTALIS_PRA74_062511,HELBIG_PRB77_245106,HEATONBURGESS_PRL98_036403}
and extensions to time-dependent CDFT have been considered~\cite{XU_PRA31_2682,DHARA_PRA35_442,GHOSH_PRA38_1149,VIGNALE_PRL77_2037,TELNOV_PRA58_4749,ULLRICH_PRB65_245102,FAASEN_PRL88_186401}.

However, in CDFT based on the charge and paramagnetic current densities as basic variables, no
HK-type theorem exists and the consequences of this have
been extensively discussed in the literature~\cite{CAPELLE_PRB65_113106}.  In the present work, we
examine this question for CDFT in some detail, demonstrating how
convex analysis of the underlying universal density functional can be
a significant aid in clarifying the relationship between basic variables
of CDFT and the potentials.
A CDFT featuring the gauge-invariant
physical current density (rather than the paramagnetic current density) as a basic variable is appealing from a
physical perspective and is therefore also considered here. Specifically, we
examine the formulations due to Diener~\cite{DIENER_JP3_9417} and Pan and
Sahni~\cite{PAN_IJQC110_2833}.

We begin in Sec.~\ref{sec:prelim} by introducing notation related
to sets of basic potentials, basic densities, and mappings between
them. In Sec.~\ref{sec:cdft}, we consider CDFTs that use the
charge and paramagnetic current densities as basic variables---in
particular, Sec.~\ref{sec:restoreconc} establishes the concavity of
a universal density functional based on these variables and
Sec.~\ref{subsec:numF} outlines the opportunities that this formulation affords for
numerical studies of this functional.  Next, in
Sec.~\ref{secPHYSCDFT}, the use of the charge and physical current
densities as basic variables is considered and two previous
formulations~\cite{PAN_IJQC110_2833,DIENER_JP3_9417} are examined.
Our concluding remarks are presented in Sec.~\ref{sec:conc}.

\section{A review of DFT} \label{sec:prelim}

Before discussing CDFT, we briefly review standard DFT, with emphasis on Lieb's treatment based on convex conjugation~\cite{LIEB_IJQC24_243}.
The concepts and techniques of convex analysis introduced here are well suited to the study of DFT and will later be used in our discussion of CDFT.
Some background is also given in the Appendix.

We consider a system of $N$ electrons with an electronic Hamiltonian of the form (in atomic units)
\begin{equation}
  H[v] = \frac{1}{2} \sum_k p_k^2 + \sum_k v(\mathbf{r}_k) + W, \label{HAMv}
\end{equation}
where $\mathbf{p}_k = - \mathrm{i} \nabla_k$ is the canonical momentum operator of electron $k$,
$v(\mathbf{r})$ is the external potential at position $\mathbf{r}$, and $W = \sum_{k<l} r_{kl}^{-1}$ is the two-electron Coulomb repulsion
operator. The state of the system is described by a density matrix $\Gamma$, which is a convex combination of normalized $N$-electron
pure-state density matrices
\begin{equation}
\Gamma = \sum_i \lambda_i \ket{\psi_i} \bra{\psi_i}, \quad \lambda_i \geq 0, \quad \sum_i \lambda_i = 1,
\label{Gamma}
\end{equation}
where the wave functions $\psi_i$ are antisymmetric in the space and spin coordinates 
$\mathbf{x}_k = (\mathbf{r}_k, \sigma_k)$ of the $N$ electrons.
The electron density associated with such density matrices is given by
\begin{equation}
\rho(\mathbf{r}) = \sum_i \lambda_i \rho_i(\mathbf{r}), \; \rho_i(\mathbf{r}_1) = N \!\int  \!\!\psi_i^{\ast} \psi_i \mathrm{d}\tau_{-1},
\label{rho}
\end{equation}
where the volume element is $\mathrm{d}\tau_{-1} = \mathrm{d}\sigma_1
\mathrm{d}\mathbf{x}_2\cdots \mathrm{d}\mathbf{x}_N$, i.e., the integration
is over all $N$ spin and spatial coordinates except $\mathbf{r}_1$.
The ground-state energy is obtained from the Rayleigh--Ritz variation
principle,
\begin{equation}
  \label{eqRR}
  E[v] = \inf_{\Gamma}  \trace{ \Gamma H[v] },
\end{equation}
where the minimization is over all $N$-electron density matrices.

An infimum rather than a minimum is taken in Eq.~\eqref{eqRR} since
$v$ may or may not support an $N$-electron ground state. The set of
potentials that support one or more $N$-electron ground states (and
for which therefore the infimum is attained) is denoted by
$\mathcal{V}_N$; the potentials in $\mathcal{V}_N$ are sometimes said
to be $\rho$-representable. Conversely, a density that is an ensemble
ground-state density for some potential $v \in \mathcal{V}_N$ is said
to be (ensemble) $v$-representable; the set of $v$-representable
densities is denoted by $\mathcal{B}_N$. For convenience, we shall also
refer to $\rho$-representable potentials and $v$-representable
densities as ground-state potentials and densities, respectively.

In the constrained-search formalism of DFT, we write the
Rayleigh--Ritz variation principle as an HK variation
principle,
\begin{equation}
E[v] = \inf_{\rho \in \mathcal{I}_N} \left( F[\rho] + ( \rho | v ) \right),
\label{eqHK}
\end{equation}
where $\mathcal{I}_N$ is the set of $N$-representable
densities---that is, the set of the nonnegative densities
$\rho$ with $\int\! \rho(\mathbf r) \mathrm d\mathbf{r} = N$ and with a finite von Weizs\"acker
kinetic energy.  The Lieb constrained-search functional $F$ is given
by
\begin{equation}
  F[\rho] = \inf_{\Gamma\mapsto \rho} \trace{\Gamma H[0]},
\label{FLL}
\end{equation}
where the notation $\Gamma \mapsto \rho$ indicates that the minimization is restricted to density matrices $\Gamma$ that reproduce the density $\rho$. If $\rho$ is not $N$-representable, no such $\Gamma$ exists, and $F[\rho]=+\infty$ by definition.
The HK variation principle
in~Eq.~\eqref{eqHK} is well defined for all potentials $v$ that have a finite pairing with every $\rho\in\mathcal{I}_N$,
\begin{equation}
  (\rho|v) = \int  \!\! \rho(\mathbf{r}) \, v(\mathbf{r}) \mathrm{d}\mathbf{r}.
  \label{eqPAIRING1}
\end{equation}

An especially attractive formulation of DFT is Lieb's formulation in terms of Legendre--Fenchel transformations or convex conjugation. This formulation is not only elegant, but also fits naturally in the well-developed mathematical field of convex analysis, 
allowing application of deep results of convex analysis to DFT, of which we will give some examples.

Lieb's formulation of DFT begins with the observation that the ground-state
energy $E[v]$ is upper semi-continuous and concave in $v$ and
therefore may be represented by its conjugate function: Lieb's
universal density functional $F[\rho]$. The ground-state energy and
density functionals are then related as
\begin{subequations}
\begin{align}
  E[v] & = \inf_{\rho\in X} \left[F[\rho] + (\rho|v)\right],
  \label{eqLIEBVFUN}
       \\
  F[\rho] & = \sup_{v\in X^*} \left[E[v] - (\rho|v)\right].
  \label{eqLIEBFFUN}
\end{align}
\label{eqLIEB}%
\end{subequations}
Here, $X$ is a Banach space (a complete normed vector space) that contains $\mathcal{I}_N$, and $X^*$ is its dual---that is, 
the set of all bounded linear functionals on $X$ (also a Banach space), 
thereby ensuring that $\vert (\rho|v) \vert<+\infty$. 
Lieb identified $X=L^1\cap L^3\subset \mathcal{I}_N$ and $X^* = L^{3/2} + L^\infty$, 
which contains, among others, all Coulomb potentials.
The observations that $E[v]$ and $F[\rho]$ are upper and lower semicontinuous concave and convex functions,
respectively, together with the identification of
the Banach spaces $X$ and $X^*$ are the key elements that place DFT
within the setting of convex analysis.

The duality of $E$ and $F$ apparent in Eq.~\eqref{eqLIEB} means
that the same information is contained in either functional but encoded in different ways; this duality is emphasized by referring to $E$ and $F$ as
the extrinsic and intrinsic energies, respectively, of the electronic system.
The infimum and supremum expressions on the right-hand sides of Eqs.~\eqref{eqLIEBVFUN} and~\eqref{eqLIEBFFUN} feature
linear pairings of densities and potentials and are therefore by construction concave and
convex, respectively.
Indeed, a necessary and sufficient condition for $E[v]$ ($F[\rho]$) to
be upper (lower) semicontinuous and concave (convex) is the existence
of such expressions~\cite{VANTIEL}.

Unlike the Lieb density-matrix constrained-search functional, the Levy--Lieb constrained-search functional, defined in terms of pure states rather than density matrices, is not convex and
hence not identical to the Lieb functional.

In the present paper, we generalize Lieb's formulation
of DFT to CDFT. In particular, we discuss the Vignale--Rasolt constrained-search functional as a generalization
of the Lieb functional to systems in the presence of a vector potential. We shall see that such a generalization is possible after a redefinition of the scalar potential.
However, we here leave aside technical questions such as lower semicontinuity and infimum and
supremum domains (i.e., Banach spaces), which is the subject of future work; for a discussion of such
mathematical issues within standard DFT, see Lieb~\cite{LIEB_IJQC24_243} and Eschrig~\cite{ESCHRIG03}.
By contrast, the concavity of $E$ and the convexity of $F$ are essential properties for our discussion of CDFT. 
If $E[v]$ happens to be non-concave, it cannot be
represented by an expression like that in Eq.~\eqref{eqLIEBVFUN}, even by allowing
$F$ to be non-convex. Therefore, no universal functional with a linear
potential pairing can exist for non-concave energies (such as those of excited states of the same symmetry as the ground state).

\section{The paramagnetic current density as a basic variable}\label{sec:cdft}

A CDFT with the paramagnetic current as a basic density was considered
in the seminal work of Vignale and
Rasolt~\cite{VIGNALE_PRL59_2360,VIGNALE_PRB37_10685}. In their
formulation of CDFT, the basic potentials are the standard
electromagnetic potentials $(v,\mathbf{A})$ and the basic densities are
the charge density and paramagnetic current density
$(\rho,\mathbf{j}_{\text{p}})$. We shall here first review their theory and then
discuss an alternative formalism, based on a redefinition of the basic
scalar potential.

\subsection{Preliminaries}

We consider electrons subject to
time-independent external electromagnetic fields
$\mathbf{E}(\mathbf{r}) = - \boldsymbol{\nabla} v(\mathbf{r})$ and
$\mathbf{B}(\mathbf{r}) = \boldsymbol{\nabla} \times \mathbf{A}(\mathbf{r})$, represented by the scalar potential
$v(\mathbf{r})$ and the vector potential $\mathbf{A}(\mathbf{r})$, respectively. For
potentials $(v, \mathbf{A})$, we introduce the equivalence relation
\begin{equation}
  (v',\mathbf{A}') \gaugeeq (v,\mathbf{A})  \iff  (\nabla v',\boldsymbol{\nabla}\times\mathbf{A}') = (\nabla v, \boldsymbol{\nabla}\times\mathbf{A}),
 \label{vAgauge}
\end{equation}
which defines equivalence classes of potentials that differ only by a
static gauge transformation, thereby representing the same external
fields.

We note that a general gauge transformation of $v$ and $\mathbf{A}$ is
given by $v' = v - \partial f/\partial t$ and $\mathbf{A}'
= \mathbf{A} + \nabla f$, for some arbitrary gauge function
$f(\mathbf{r}, t)$. If $\mathbf{A}$ is to remain static after the
transformation, we must require that $f(\mathbf{r},t) = \chi(\mathbf{r}) -
c t$, where $c$ is constant. It follows that a general
time-independent gauge transformation is given by $v' = v + c$ and
$\mathbf{A}' = \mathbf{A} + \boldsymbol{\nabla} \chi$, where the
constant $c$ and the function $\chi(\mathbf{r})$ are independent.
Therefore, the equivalence relation in Eq.~\eqref{vAgauge} holds if
and only if there exists a constant $c$ and a sufficiently
well-behaved gauge function $\chi(\mathbf{r})$ such that $v' = v+c$
and $\mathbf{A}' = \mathbf{A} + \boldsymbol{\nabla}\chi$.

In the presence of a vector potential, the electronic Hamiltonian in
Eq.~\eqref{HAMv} is modified by replacing the canonical momentum
operator $\mathbf{p}_k = - \mathrm{i}\nabla_k$ by the mechanical
(kinetic) momentum operator $\boldsymbol{\pi}_k = -\mathrm{i}\nabla_k
+ \mathbf{A}(\mathbf{r}_k)$, yielding
\begin{equation}
  H[v,\mathbf{A}] = \frac{1}{2} \sum_k \pi_k^2 + \sum_k v(\mathbf{r}_k) + W.
  \label{HAM}
\end{equation}
We have here omitted the spin-dependent term,
$\sum_k \mathbf{B}(\mathbf{r}_k)\cdot\mathbf{S}$, from the Hamiltonian.
By analogy with Eq.~\eqref{eqRR}, the Rayleigh--Ritz variation principle in the presence of a vector potential is given by
\begin{equation}
  \label{eqCDFTVFUN}
  E[v,\mathbf{A}] = \inf_{\Gamma}  \trace{ \Gamma H[v,\mathbf{A}] },
\end{equation}
where the minimization is over density matrices $\Gamma$ containing $N$ electrons, see Eq.~\eqref{Gamma}. An infimum rather than
a minimum is taken to ensure that the energy is well defined also when $(v,\mathbf{A})$ does not support a ground state.
We denote the set of all potentials $(v,\mathbf A)$ that support a ground state with this Hamiltonian by
\begin{equation}
  \mathcal{V}_N = \{(v,\mathbf{A}) | H[v,\mathbf{A}] \text{\ has a g.s.}\}
\end{equation}
and also introduce the related set 
\begin{equation}
  \mathcal{U}_N = \{(v+\tfrac{1}{2} A^2,\mathbf{A}) | H[v,\mathbf{A}] \text{\ has a g.s.}\}
\end{equation}
in preparation of a reparameterization of the scalar potential that will be introduced later.

In the presence of a vector potential, the ensemble ground-state charge densities $\rho$ are as before given by Eq.~\eqref{rho}.
Regarding the induced currents, we distinguish between
the paramagnetic current density and the physical current density. The
former is defined as
\begin{equation}
\mathbf{j}_{\text{p}}(\mathbf{r}_1) = \mathrm{Re} \sum_{ik} \lambda_i \int  \! \psi_i^\ast \mathbf{p}_k \psi_i \, \mathrm{d}\tau_{-1}.
\end{equation}
The paramagnetic current density is gauge-dependent and unobservable. 
The physical current density is given by
\begin{equation}
\mathbf{j}(\mathbf{r}_1) = \mathrm{Re} \sum_{ik} \lambda_i \int \! \psi_i^* \boldsymbol{\pi}_k \psi_i \mathrm{d}\tau_{-1},
\end{equation}
and satisfies the relation $\mathbf{j} = \mathbf{j}_{\text{p}} + \rho
\mathbf{A}$. 
Unlike the paramagnetic current, the physical current is gauge-invariant. 

Finally, the sets of paramagnetic and physical $v$-representable ground-state densities are denoted by 
\begin{align}
  \mathcal{B}^{\text{p}}_N &= \{(\rho,\mathbf{j}_{\text{p}}) | (\rho,\mathbf{j}_{\text{p}}) \text{\ is g.s. den.\ of some\ } H[v,\mathbf{A}]\}, \\
  \mathcal{B}_N &= \{(\rho,\mathbf{j}) | (\rho,\mathbf{j}) \text{\ is g.s. den.\ of some\ } H[v,\mathbf{A}]\},
\end{align}
where both mixed and pure states are allowed.

\subsection{Do paramagnetic densities determine potentials?}

The HK theorem of standard DFT states that the ground-state 
density $\rho$ determines the scalar potential $v$ up to a
constant shift. Hence, two potentials that differ by more than a
constant shift cannot give rise to the same ground-state density. This
fact establishes a mapping from ground-state densities to potentials.

Vignale and Rasolt established that two
different potentials $(v_1,\mathbf{A}_1) \neq (v_2,\mathbf{A}_2)$ with different ground-state wave functions $\psi_1 \neq \psi_2$ cannot
give rise to the same paramagnetic ground-state density $(\rho,\mathbf{j}_{\text{p}})$. 
However, this does not establish an analogue of the HK
theorem for CDFT~\cite{CAPELLE_PRB65_113106} since different
potentials $(v_1,\mathbf{A}_1) \neq (v_2,\mathbf{A}_2)$ can map to the same density in
$\mathcal{B}^{\text{p}}_N$ via the same wave function,
$\Psi[v_1,\mathbf{A}_1] = \Psi[v_2,\mathbf{A}_2]$. Vignale and
Rasolt's result can be extended to a form that applies also when the
ground states are degenerate: {\it Let $\psi_1$ be a
  ground state of $H[v_1,\mathbf{A}_1]$ and let $\psi_2$ be a ground
  state of $H[v_2,\mathbf{A}_2]$. If $\psi_1$ and $\psi_2$ give rise
  to the same paramagnetic density,
  $(\rho_{\psi_1},\mathbf{j}_{\text{p};\psi_1}) =
  (\rho_{\psi_2},\mathbf{j}_{\text{p};\psi_2})$, then $\psi_2$ is also
  a ground state of $H[v_1,\mathbf{A}_1]$ and $\psi_1$ is also a
  ground state of $H[v_2,\mathbf{A}_2]$.}

The above statement appears to be as close as one can get to a
HK-like result for paramagnetic densities
$(\rho,\mathbf{j}_{\text{p}})$. A CDFT formulated in terms of the
paramagnetic current density thus cannot be based on a formal mapping
from ground-state densities to potentials. On the other hand, rigorous
formulations of standard DFT do not rely on the HK
mapping from densities to potentials. These rigorous formulations can
be extended to the paramagnetic current density as a basic variable.
The absence of an HK-type theorem in CDFT is therefore not
a serious impediment.


\subsection{The standard electromagnetic potentials as basic potentials}
\label{sec:electronic}

An important property of $E[v]$ in Eq.~\eqref{eqLIEBVFUN} is its
concavity in $v$, which established duality with
the universal density functional $F[\rho]$.  However, unlike $E[v]$,
the energy functional $E[v,\mathbf{A}]$ in Eq.~\eqref{eqCDFTVFUN} is
not concave.  The non-concavity of $E[v,\mathbf{A}]$ is apparent in,
for example, any diamagnetic ground state at vanishing external
magnetic field.  Such a ground state has a negative definite
magnetizability tensor $\boldsymbol{\chi}$ and, when restricted to
weak uniform magnetic fields $\mathbf{B}\approx \mathbf{0}$, the
energy is a convex function $E_0 - \frac{1}{2} \mathbf{B}^T
\boldsymbol{\chi} \mathbf{B}$ in $\mathbf{B}$ and therefore in
$\mathbf{A}$.

In more detail, consider a one-electron system confined to the
(two-dimensional) $xy$-plane, subject both to a uniform magnetic field
along the $z$-axis and to a harmonic-oscillator
potential. Parameterizing the scalar and vector potentials under
consideration as
\begin{equation}
  v_{\text{HO}}(\mathbf{r};k) = \frac{1}{2} k (x^2 + y^2), \quad \mathbf{A}_{\perp}(\mathbf{r};B) = \frac{1}{2} B \mathbf{e}_z\times\mathbf{r},
\end{equation}
we obtain the following Hamiltonian
\begin{equation}
 \begin{split}
  H[v_{\text{HO}},\mathbf{A}_{\perp}] & = \frac{1}{2} p^2 + \frac{1}{2} B L_z + v_{\text{HO}}(\mathbf{r};k) + \frac{1}{2} A_{\perp}(\mathbf{r};B)^2 \\
  & = \frac{1}{2} p^2 + \frac{1}{2} B L_z + \frac{1}{8} \left(4k + B^2\right) (x^2+y^2),
 \end{split}
\end{equation}
where $L_z$ is a good quantum number. For $k \geq k_{\text{min}} > 0$ and
some finite interval $|B| < B_{\text{max}}$, the ground state has $L_z
= 0$ and application of a magnetic field has exactly the same effect as the introduction of a
harmonic-oscillator potential. For these potentials, the ground-state energy is
\begin{equation}
  E[v_{\text{HO}},\mathbf{A}_{\perp}] = E[v_{\text{HO}} + \tfrac{1}{2}
  A_{\perp}^2, \mathbf{0}] = \sqrt{k + \tfrac{1}{4} B^2}.
  \label{eq:example-energy}
\end{equation}
Note that the right-hand side is concave in $k\geq 0$ and convex in
$B$. Hence, on the restricted set of potentials spanned by $k\geq
k_{\text{min}}$ and $|B| < B_{\text{max}}$, the functional
$E[v_{\text{HO}},\mathbf{A}_{\perp}]$ is not only non-concave but
convex in its second argument. On a larger domain, the functional is
neither concave nor convex.  We conclude that $E[v,\mathbf A]$ cannot
be represented by a conjugate functional in the manner of
Eq.~\eqref{eqLIEB}.  However, this does not preclude a
constrained-search formulation of CDFT, as discussed in the next
subsection.

\subsection{CDFT by constrained search}

Rewriting the Rayleigh--Ritz variation principle in Eq.~\eqref{eqCDFTVFUN}
by analogy with the constrained-search approach of standard DFT in Eq.~\eqref{FLL}, we obtain an HK-type variation principle for a system in the presence
of a scalar and vector potential,
\begin{equation}
  \label{eqCDFTVFUN2}
  E[v,\mathbf{A}] = \inf_{\rho,\mathbf{j}_{\text{p}}} \left[ F_{\text{VR}}[\rho,\mathbf{j}_{\text{p}}] + (\rho|v+\tfrac{1}{2} A^2) + (\mathbf{j}_{\text{p}}|\mathbf{A}) \right],
\end{equation}
where the \emph{Vignale--Rasolt constrained-search functional} is given by
\begin{equation}
  F_{\text{VR}}[\rho,\mathbf{j}_{\text{p}}] = \inf_{\Gamma\mapsto(\rho,\mathbf{j}_{\text{p}})} \trace{\Gamma (\tfrac{1}{2} p^2 + W)}
\label{FVR}
\end{equation}
and we have introduced the following notation for the pairing between a current density and a vector potential: 
\begin{align}
  (\mathbf{j}_{\text{p}}|\mathbf{A}) &= \int  \mathbf{j}_{\text{p}}(\mathbf{r}) \cdot  \mathbf{A}(\mathbf{r}) \, \mathrm{d}\mathbf{r}.
\label{eqPAIRING2}
\end{align}

Like the Lieb constrained-search functional given in Eq.~\eqref{FVR}, the
Vignale--Rasolt constrained-search functional in~Eq.~\eqref{FVR} is universal in the sense that it does not depend on
the potential $(v,\mathbf{A})$, only on the density $(\rho,\mathbf{j}_{\text{p}})$. Another important characterization of the Lieb functional
is its convexity. To examine the convexity of $F_{\text{VR}}$, let $(\rho_1,\mathbf{j}_{\text{p}1})$ and
$(\rho_2,\mathbf{j}_{\text{p}2})$ be arbitrary (Lebesque integrable)
functions and let $0<\lambda<1$, $\mu=1-\lambda$. We then obtain
\begin{equation}
\begin{split}
  F_{\text{VR}}&[\lambda\rho_1  + \mu\rho_2, \lambda
    \mathbf{j}_{\text{p}1} + \mu\mathbf{j}_{\text{p}2}] \\
    & \leq
  \inf_{\substack{\Gamma_1\mapsto(\rho_1,\mathbf{j}_{\text{p}1})\\\Gamma_2\mapsto(\rho_2,\mathbf{j}_{\text{p}2})}}
      \trace{(\lambda\Gamma_1 + \mu\Gamma_2)  (\tfrac{1}{2} p^2 + W)}  \\
     & = \lambda F_{\text{VR}}[\rho_1 , \mathbf{j}_{\text{p}1} ] +
      \mu F_{\text{VR}}[\rho_2,\mathbf{j}_{\text{p}2}],
\end{split}
\end{equation}
demonstrating that $F_{\text{VR}}$ is convex in $(\rho,\mathbf{j}_{\text{p}})$.
The key point in establishing the inequality above is
to restrict the infimum over all density matrices $\Gamma \mapsto (\rho,\mathbf{j}_{\text{p}})$ to an infimum over all matrices of the form
$\Gamma = \lambda\Gamma_1 + \mu\Gamma_2$ where
$\Gamma_1 \mapsto (\rho_1,\mathbf{j}_{\text{p}1})$ and
$\Gamma_2 \mapsto (\rho_2,\mathbf{j}_{\text{p}2})$, thereby overestimating the infimum.

Given that the Vignale--Rasolt functional is convex, it is uniquely represented by a convex conjugate functional.
For the Lieb functional $F[\rho]$, the conjugate is the concave ground-state energy $E[v]$. However, since $E[v,\mathbf{A}]$
is not concave, it cannot be the conjugate to $F_{\text{VR}}[\rho,\mathbf j_\text p]$. In the following, we identify the energy conjugate
to $F_\text{VR}[\rho,\mathbf j_\text p]$, thereby arriving at a Legendre--Fenchel formulation of CDFT.

\subsection{CDFT by convex conjugation}
\label{sec:restoreconc}

Inspection of the general expression in Eq.~\eqref{eqCDFTVFUN2}
and the harmonic-oscillator example suggest the introduction of a new
basic scalar potential,
\begin{equation}
  u = v + \tfrac{1}{2} A^2.
\end{equation}
The choice of $(u,\mathbf{A})$ as basic potentials and
$(\rho,\mathbf{j}_{\text{p}})$ as basic densities results in a theory
where the HK variation principle takes the form of a
Legendre--Fenchel transformation with a linear pairing 
$(\rho|u) + (\mathbf{j}_{\text{p}}|\mathbf{A})$ of the densities and potentials:
\begin{equation}
 \label{eqCDFTEUFUN}
 \begin{split}
  \bar{E}[u,\mathbf{A}] & = \inf_{\Gamma} \trace{\Gamma \bar{H}[u, \mathbf{A}]} \\
   & = \inf_{\rho,\mathbf{j}_{\text{p}}} \left[F_{\text{VR}}[\rho,\mathbf{j}_{\text{p}}] + (\rho|u) + (\mathbf{j}_{\text{p}}|\mathbf{A}) \right].
 \end{split}
\end{equation}
Here we have introduced the notation
\begin{align}
\bar{H}[u,\mathbf{A}] &= H[u - \tfrac{1}{2} A^2, \mathbf{A}], \\
\bar{E}[u,\mathbf{A}] &= E[u - \tfrac{1}{2} A^2, \mathbf{A}] .
\end{align}
The energy $\bar{E}[u,\mathbf{A}]$ is now by construction concave,
allowing it to be generated from the convex intrinsic energy
$F_{\text{VR}}[\rho,\mathbf{j}_{\text{p}}]$ by a reverse Legendre--Fenchel transformation:
\begin{equation}
  \label{eqFVRLEGENDRE}
  F_{\text{VR}}[\rho,\mathbf{j}_{\text{p}}] = \sup_{u,\mathbf{A}} \left[ \bar{E}[u,\mathbf{A}] - (\rho|u) - (\mathbf{j}_{\text{p}}|\mathbf{A}) \right].
\end{equation}
Thus, by a change of variables from $v$ to $u=v + \tfrac{1}{2} A^2$,
we have restored the conjugate relation between the extrinsic and intrinsic energies characteristic of standard DFT.

Strictly speaking, for $F_{\text{VR}}[\rho,\mathbf{j}_{\text{p}}]$ and
$\bar{E}[u,\mathbf{A}]$ to form a conjugate pair, we must specify
their domains in the form of a Banach space $X_{\text{p}}$ and its
dual $X_{\text{p}}^*$, respectively. Furthermore, we must demonstrate lower and
upper semi-continuity of the intrinsic and extrinsic
energies, respectively.  
However, regarding the domains, we note here that, in Lieb's formulation of
standard DFT, the vector space $X^* = L^{3/2} + L^\infty$ of potentials does not contain
every potential with a square integrable ground state (e.g., it does
not contain harmonic potentials), but $X^\ast$ does contain all
Coulomb potentials, thus covering most systems of interest.  A similar
compromise is expected for CDFT: we cannot expect to
identify a vector space $X_{\text{p}}$ of densities such that its dual
$X_{\text{p}}^*$ includes all ground-state potentials $\mathcal{V}_N$.
In particular, since gauge transformations may produce potentials
$\mathbf{A}+\boldsymbol{\nabla}\chi$ that are arbitrarily ill-behaved
at infinity, we expect some gauge restriction to be necessary.

\subsection{Subdifferentiability in CDFT}
\label{sec:subdiff-cdft}

In general, the functionals $\bar{E}$ and $F_{\text{VR}}$ are not differentiable. Consequently, 
we cannot characterize the ground-state densities in the CDFT HK
variation principle of Eq.~\eqref{eqCDFTEUFUN} in terms of functional derivatives. On the other hand, in convex analysis, the proper tool for characterizing
minimizers and maximizers are sub- and supergradients, respectively. We here introduce sub- and supergradients in the context of CDFT.

\begin{figure}
  \includegraphics{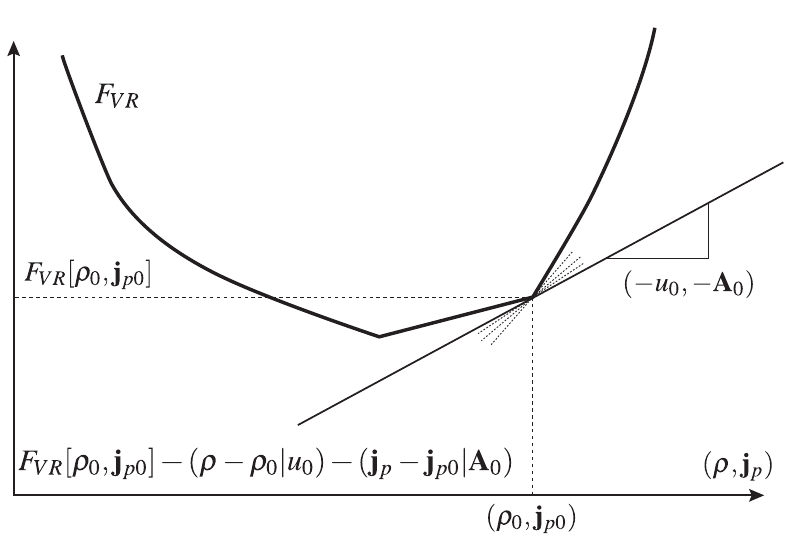}
  \caption{Schematic illustration of the subdifferential
    $\subdiff{F}_{\text{VR}}[\rho_0,\mathbf{j}_{\text{p}0}]$: The set of
    all slopes (potentials $-(u,\mathbf{A})$) of just-touching tangent
    planes entirely below the graph at $(\rho_0,\mathbf{j}_{\text{p}0})$. One
    particular tangent, with slope $-(u_0,\mathbf{A}_0)$, is shown, while
    others are indicated by dashed lines. A similar illustration can be made of the
    superdifferential
    $\supdiff \bar{E}[u,\mathbf{A}]$.
    \label{fig:subdiff}}
\end{figure}

We begin by noting that an immediate consequence of the CDFT variation principles in Eqs.~\eqref{eqCDFTEUFUN} and~\eqref{eqFVRLEGENDRE} is
\emph{Fenchel's inequality},
\begin{equation}
  \bar{E}[u,\mathbf{A}] \leq
  F_{\text{VR}}[\rho,\mathbf{j}_{\text{p}}] + (\rho|u) + (\mathbf{j}_{\text{p}}|\mathbf{A}),
  \label{eqFENCHELINEQ}
\end{equation}
valid for any choice of potential $(v,\mathbf{A})\in X_{\text{p}}^*$ and density $(\rho,\mathbf{j}_{\text{p}})\in X_{\text{p}}$.
Moreover, equality holds if and only if $(\rho,\mathbf{j}_{\text{p}}) \in \mathcal{B}_N^{\text{p}}$ is a ground-state density belonging to $(u,\mathbf{A})$.
To characterize ground-state densities and their potentials mathematically, we use the concepts of subgradients and subdifferentials. 
A \emph{subgradient} of $F_{\text{VR}}$ at $(\rho_0,\mathbf{j}_{\text{p}0}) \in X_{\text{p}}$ is an external potential
$-(u_0,\mathbf{A}_0)\in X_{\text{p}}^*$ for which the inequality
\begin{equation}
  F_{\text{VR}}[\rho,\mathbf{j}_{\text{p}}] \geq F_{\text{VR}}[\rho_0,\mathbf{j}_{\text{p}0}] - (\rho-\rho_0|u_0) - (\mathbf{j}_{\text{p}}-\mathbf{j}_{\text{p}0}|\mathbf{A}_0)
\end{equation}
holds for all $(\rho,\mathbf{j}_{\text{p}})\in X_{\text{p}}$; see Fig.\,\ref{fig:subdiff}. Clearly, all potentials
$-(u_0,\mathbf{A}_0)$ for which the density $(\rho_0,\mathbf{j}_{\text{p}0})$ is a minimizer in
Eq.~\eqref{eqCDFTEUFUN} are subgradients of $F_\text{VR}$ at $(\rho_0,\mathbf{j}_{\text{p}0})$.
The set of all subgradients at $(\rho_0,\mathbf{j}_{\text{p}0})$ is known as the
\emph{subdifferential} of $F_{\text{VR}}$ at $(\rho_0,\mathbf{j}_{\text{p} 0})$ and is denoted by $\subdiff{F}_{\text{VR}}[\rho_0,\mathbf{j}_{\text{p}0}]\subset X_{\text{p}}^*$.
Hence, to within a minus sign, the subdifferential at $(\rho_0,\mathbf j_{\text p 0})$ 
is the collection of all external potentials that have the same ground-state density $(\rho_0,\mathbf j_{\text p 0})$.

Analogously, we consider the concave energy functional $\bar{E}[u,\mathbf{A}]$ and its supergradients.
In general, $(\rho_0,\mathbf{j}_{\text{p}0})\in X_{\text{p}}$ is a
\emph{supergradient} of $\bar{E}$ at $(u_0,\mathbf{A}_0) \subset X_{\text{p}}^\ast$ if and only if
$-(\rho_0,\mathbf{j}_{\text{p}0})$ is a subgradient of the convex functional
$-\bar{E}$ at $(u_0,\mathbf{A}_0)$. Hence, $(\rho_0,\mathbf{j}_{\text{p}0})$ is a
supergradient of $\bar{E}$ at $(u_0,\mathbf{A}_0)$ if the inequality
\begin{equation}
  \bar{E}[u,\mathbf{A}] \leq \bar{E}[u_0,\mathbf{A}_0] + (\rho_0|u-u_0) + (\mathbf{j}_{\text{p}0}|\mathbf{A}-\mathbf{A}_0)
\end{equation}
holds for all $(u,\mathbf{A})\in X_{\text{p}}^*$. This condition is satisfied precisely
when $(\rho_0,\mathbf{j}_{\text{p}0})$ is the density arising from a (possibly
degenerate) ground state of $\bar{H}[u_0,\mathbf{A}_0]$. The
\emph{superdifferential} $\supdiff \bar{E}[u_0,\mathbf{A}_0]$ is
the collection of all supergradients of $\bar{E}$ at $(u_0,\mathbf{A}_0)$.

\begin{figure}
  \includegraphics{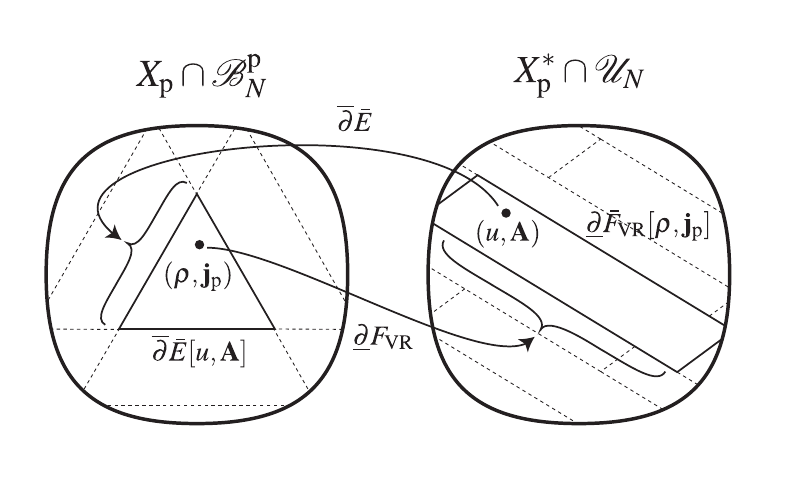}
  \caption{Illustration of considerations in
    Sec.~\ref{sec:subdiff-cdft}. Instead of a one-to-one mapping of
    individual potential pairs and corresponding ground-state density
    pairs, the sub- and superdifferentials of $F_{\text{VR}}$ and
    $\bar{E}$, respectively, maps (ground-state) densities into convex
    sets of (ground-state) potentials and potentials into convex sets
    of densities.\label{fig:subdiff2}}
\end{figure}

For all ground-state densities $(\rho_0,\mathbf{j}_{\text{p}0})$ and associated potentials
$(u_0,\mathbf{A}_0)$, we now have the following stationary conditions of the HK and Lieb variation principles in 
Eqs.~\eqref{eqCDFTEUFUN} and~\eqref{eqFVRLEGENDRE}, respectively:
\begin{alignat}{2}
  -(u_0, \mathbf{A}_0) &\in \subdiff{F}_{\text{VR}}[\rho_0,\mathbf{j}_{\text{p}0}], \quad & (\rho_0,\mathbf{j}_{\text{p}0}) &\in \mathcal{B}_N^{\text{p}} \cap X,\\
  (\rho_0,\mathbf{j}_{\text{p}0}) &\in \supdiff \bar{E}[u_0,\mathbf{A}_0],  \quad & (u_0,\mathbf{A}_{0}) &\in \mathcal{U}_N \cap X^\ast.
\end{alignat}%
Importantly, these conditions are equivalent: $-(u_0,\mathbf{A}_0)$ is a subgradient of $F_{\text{VR}}$ at $(\rho_0,\mathbf{j}_{\text{p}0})$ if and only if
$(\rho_0,\mathbf{j}_{\text{p}0})$ is a supergradient of $\bar{E}$ at $(u_0,\mathbf{A}_0)$. Hence,
instead of a one-to-one mapping between individual potentials and
individual ground-state densities, the convexity and concavity of the intrinsic and extrinsic energies, respectively,
establish a mapping between the convex sets
$\{(\rho,\mathbf{j}_{\text{p}})\}\subset X_{\text{p}}$ of degenerate ground-state densities and
convex sets $\{(u,\mathbf{A})\}\subset X_{\text{p}}^*$ of potentials that give rise to
identical ground-state densities; see Fig.~\ref{fig:subdiff2}.

The sub- and super\-differentials are empty when no minimizer and maximizer exist in the corresponding optimization problems in
Eqs.~\eqref{eqCDFTEUFUN} and~\eqref{eqFVRLEGENDRE}. However, it is a general result of convex analysis that
the subgradients (supergradients) of a convex (concave) function (under certain semicontinuity conditions) exist at a dense subset of the domain of the function.
In CDFT, this
result implies that the set of ground-state densities $(\rho_0, \mathbf{j}_{\text{p},0})$ is dense in the set of all densities
and that the set of potentials $(u_0,\mathbf{A}_0)$ that support a ground state is dense in the set of potentials.

\subsection{Degeneracies in CDFT}
\label{sec:subdiff-cdft-deg}

\begin{figure}
  \includegraphics{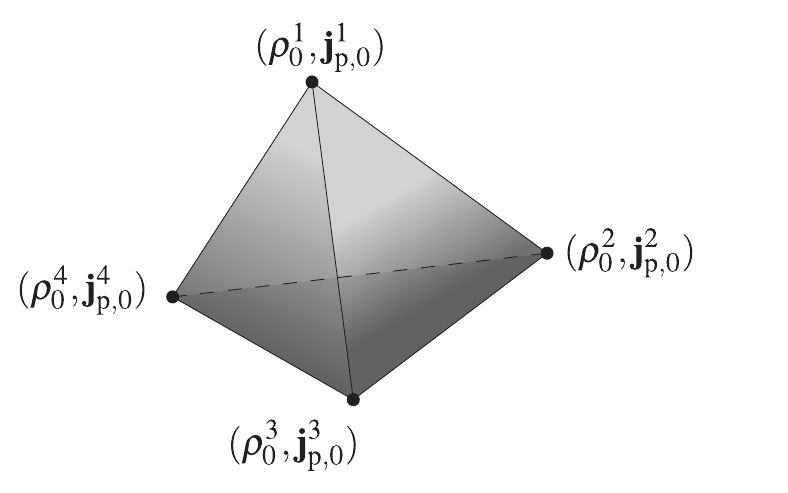}
  \caption{Illustration of the set
    $\bar{\partial}\bar{E}[u,\mathbf{A}]$ of ground state densities
    for potentials $(u,\mathbf{A})$ with degeneracy
    $G_\text{d}=4$. This is a convex set, a three-dimensional simplex,
    with vertices at $(\rho_0^i,\mathbf{j}_{\text{p},0}^i)$, i.e.,
    $\bar{\partial}\bar{E}[u,\mathbf{A}] =
    \co\{(\rho_0^i,\mathbf{j}_{\text{p},0}^i)\}.$ In this example, we
    embed the simplex in $\mathbb{R}^3$, but in reality it is a
    three-dimensional subset of the density space $X_{\text{p}}$.  The
    corresponding set of potentials is also a convex set, but usually
    with a more complicated structure. See also Fig.~\ref{fig:subdiff2}.
    \label{fig:convex-hull}
}
\end{figure}

Consider now the case of degenerate ground-state densities in
Eq.~\eqref{eqCDFTEUFUN}. For a finite degeneracy $G_\text d$, the
superdifferential of the ground-state energy in the external potential
$(u_0,\mathbf{A}_0)$ is then a $({G_\text d}-1)$-dimensional simplex
with $G_\text d$ pure-state densities $(\rho_0^i, \mathbf{j}_{\text{p}
  0}^i)$ at the vertices:
\begin{equation}
  \supdiff{\bar{E}}[u_0,\mathbf{A}_0] = \co\{(\rho_0^i, \mathbf{j}_{\text{p},0}^i), \, | \, i = 1, G_\text d\},
 \label{corho}
\end{equation}
see Fig.~\ref{fig:convex-hull} for an illustration.
Consequently, each ground-state density $(\rho_0,\mathbf{j}_{\text{p} 0})$ may be written as convex combination of the $G_{\text{d}}$ pure-state densities,
\begin{align}
(\rho_{0}, \mathbf{j}_{\text{p} 0}) = \sum_{i=1}^{G_\text d} \lambda_i (\rho_0^i, \mathbf{j}_{\text{p} 0}^i), \; \sum_{i=1}^{G_{\text{d}}} \lambda_i = 1, \; \lambda_i \geq 0.
\end{align}
As is well known, such degeneracies are either accidental or caused by symmetries of the Hamiltonian.

Consider next a ground-state density $(\rho_0,\mathbf{j}_{\text{p} 0})$ with several maximizing potentials in the Lieb variation principle in Eq.~\eqref{eqFVRLEGENDRE}.
Like the superdifferential of the ground-state energy, 
the subdifferential of the Vignale--Rasolt density functional is a convex set. Let now
$(u_0,\mathbf{A}_0)$ be a potential with ground-state density $(\rho_0, \mathbf{j}_{\text{p} 0})$ and consider the family of external potentials
\begin{equation}
(u_{0}(\lambda), \mathbf{A}_{0}(\lambda)) = (u_{0},\mathbf{A}_{0}) + \lambda (\Delta u_{0}, \Delta \mathbf{A}_{0}), \label{uAl}
\end{equation}
for some potential $(\Delta u_0,\Delta \mathbf{A}_0)$ and $\lambda \in \mathbb R$. The
associated Hamiltonians are given by
\begin{align}
\bar{H}[u_0(\lambda),\mathbf{A}_0(\lambda)] &= \bar{H}[u_0,\mathbf{A}_0] + \lambda K, \\
K &= \sum\nolimits_i [\Delta u_0(\mathbf{r}_i) + \frac{1}{2} \{\mathbf{p}_i ,\Delta \mathbf{A}_0(\mathbf{r}_i)\}] .
\end{align}
where $\{\Lambda,\Omega\}$ denotes the anti-commutator.
Clearly, if $\bar{H}[u_0,\mathbf{A}_0]$ and $K$ commute,
\begin{equation}
\left[ \bar{H}[u_0,\mathbf{A}_0], K \right] = 0, \label{HuK}
\end{equation}
then $\bar{H}[u_0(\lambda),\mathbf{A}_0(\lambda)]$ has the same
eigenstates for all $\lambda$. In particular, as $\lambda$ changes
continuously from $0$, the external potentials in Eq.~\eqref{uAl}
will have the same ground-state density
$(\rho_0,\mathbf{j}_{\text{p}0})$ until a level crossing with the
ground state occurs for $\bar{H}[u_0(\lambda),\mathbf{A}_0(\lambda)]$,
see Capelle and Vignale~\cite{CAPELLE_PRB65_113106}.  We note that the
commutation condition in Eq.~\eqref{HuK} is sufficient but not
necessary for the existence of degenerate maximizing potentials.

There may be several, independent perturbing potentials that commute with the reference Hamiltonian $\bar{H}[u_0,\mathbf{A}_0]$.
Assuming a finite degeneracy $G_{\text{p}}$ of these potentials, we may write the subdifferential of the Vignale--Rasolt density functional as a $(G_{\text{p}} - 1)$-dimensional simplex
\begin{equation}
\subdiff{F}_{\text{VR}}[\rho_0,\mathbf{j}_{\text{p}0}] = -\co\{(u_0^i, \mathbf{A}_{0}^i), \,|\, i = 1, G_{\text{p}}\}.
\end{equation}
All potentials with the ground-state density $(\rho_0, \mathbf{j}_{\text{p} 0})$ may then be written as a convex combination,
\begin{align}
(u_{0}, \mathbf{A}_{0}) = \sum_{i=1}^{G_{\text{p}}} \lambda_i (u_0^i, \mathbf{A}_{\text{p} 0}^i), \; \sum_{i=1}^{G_{\text{p}}} \lambda_i = 1, \; \lambda_i \geq 0,
\end{align}
of the $G_{\text{p}}$ vertex potentials $(u_0^i,\mathbf{A}_{0}^i)$.

Consider now the special case of DFT, for which the HK and Lieb variation principles are given by Eqs.~\eqref{eqLIEBVFUN}
and~\eqref{eqLIEBFFUN}, where $E[v]$ and $F[\rho]$ depend only on the scalar potential and the electron density, respectively.
Let $v_0$ be a potential with ground-state density $\rho_0$.
Clearly, the only scalar potentials $v(\mathbf{r})$
that commute with the Hamiltonian $H[v_0]$ are the constant potentials $c$. It follows that
the subdifferential of $F$ at $\rho_0$ is the convex set
\begin{equation}
\subdiff{F}[\rho_0] = - \{ v_0(\mathbf{r}) + c \,|\, c \in \mathbb R \},
\end{equation}
which may be regarded as a one-dimensional simplex with vertices $-v_0(\mathbf{r}) \pm \infty$. In DFT, therefore, the ground-state density $\rho_0$
determines the external potential $v_0(\mathbf{r})$ uniquely up to an additive constant $c$, in accordance with the HK theorem.

Returning to CDFT, consider next two potentials $(u_1,\mathbf{A}_1)$ and $(u_2,\mathbf{A}_2)$ with the same
ground-state density $(\rho_0,\mathbf{j}_{\text{p}0})$. By the convexity of the subgradient of $F_{\text{VR}}$, all convex combinations
\begin{equation}
(u, \mathbf{A}) = \lambda (u_1, \mathbf{A}_1) + (1-\lambda) (u_2,\mathbf{A}_2), \quad 0 \leq \lambda \leq 1 \label{NonUnique}
\end{equation}
then have the same ground-state density.
Recalling that $u_1 = v_1 + \tfrac{1}{2} A_1^2$,
$u_2 = v_2 + \tfrac{1}{2} A_2^2$, and $u = v + \tfrac{1}{2} A^2$, the
characterization of the non-uniqueness given in Eq.~\eqref{NonUnique} can be expressed in
terms of the ordinary scalar potential $v$. If $(v_1,\mathbf{A}_1)$ and
$(v_2,\mathbf{A}_2)$ give rise to the same density, then so do
all potentials of the form
\begin{equation}
 \begin{pmatrix}
    v \\ \mathbf{A}
 \end{pmatrix}
   =
    \begin{pmatrix}
      \lambda v_1 + (1-\lambda) v_2 + \tfrac{1}{2} \lambda (1-\lambda) |\mathbf{A}_1-\mathbf{A}_2|^2 \\
      \lambda \mathbf{A}_1 + (1-\lambda) \mathbf{A}_2
     \end{pmatrix} 
\end{equation}
with $0 < \lambda < 1$. However, this set is not a convex set and not a subdifferential, 
due to the use of the $(v,\mathbf{A})$ rather than $(u,\mathbf A)$ variables.

An advantage 
of the formulation of stationary conditions in CDFT in terms of sub- and superdifferentials is that differentiability is not required. 
In general, a sufficient condition for differentiability of a function at a point is that the function is continuous at this point and has
a single sub- or supergradient there; in the absence of continuity, differentiability is not guaranteed.
The ground-state
energy is differentiable at all potentials $(u,\mathbf{A})$ that have a nondegenerate ground-state density, whereas the Vignale--Rasolt functional is in principle
nowhere differentiable since we may always add a constant term $c$ to the potential without affecting the ground-state densities. However, assuming that this is
the only cause of nondifferentiability of the potentials, we may in the absence of other degeneracies write
\begin{equation}
  \frac{\delta F_{\text{VR}}[\rho,\mathbf{j}_p]}{\delta \rho(\mathbf{r})} = -u(\mathbf{r}) - c,  \quad
   \frac{\delta F_{\text{VR}}[\rho,\mathbf{j}_{\text{p}}]}{\delta \mathbf{j}_{\text{p}}(\mathbf{r})} = -\mathbf{A}(\mathbf{r}),
  \label{statF}
\end{equation}
and
\begin{equation}
  \frac{\delta \bar{E}[u,\mathbf{A}]}{\delta u(\mathbf{r})} = \rho(\mathbf{r}), \quad
  \frac{\delta \bar{E}[u,\mathbf{A}]}{\delta \mathbf{A}(\mathbf{r})} = \mathbf{j}_{\text{p}}(\mathbf{r}),
  \label{statE}
\end{equation}
where $(\rho,\mathbf{j}_{\text{p}})$ is the (non-degenerate) ground-state density of the potential $(u,\mathbf{A})$.

\subsection{Numerical calculation of $F_{\text{VR}}[\rho, \mathbf{j}_{\text{p}}]$}\label{subsec:numF}

Besides allowing formal application of theorems in convex analysis to
CDFT, the convex formulation given above has practical value. After linear
programming and optimization of quadratic functions, optimization of
convex and concave functions is the mathematically most
well-characterized type of optimization problem. The fact that convex and concave 
optimization problems have a unique global optimum (either in the form a
single point or a convex set of optimal points) and no additional
local optima is of great value when devising practical optimization
methods.

In standard DFT, Lieb's formulation~\cite{LIEB_IJQC24_243} of $F[\rho]$ in terms of the Legendre--Fenchel
transform of the concave energy functional $E[v]$ has proven useful in the study of functionals of interest
in Kohn--Sham theory~\cite{COLONNA_JCP110_2828,WU_JCP118_2498,TEALE_JCP130_104111,TEALE_JCP132_164115,TEALE_JCP133_164112,STROMSHEIM_JCP135_194109}. In particular, the modulation of the two-electron interaction
operator $W$ by a parameter $\lambda$ such that,
\begin{equation}
  W_{\lambda} = \sum_{i<j} w_\lambda(r_{ij}), \; w_0(r_{ij}) = 0, \; w_1(r_{ij}) = 1/r_{ij},
  \label{WLAM}
\end{equation}
allows us to represent the ground-state energy $E_\lambda[v]$ at interaction strength $\lambda$
in terms of its conjugate functional $F_{\lambda}[\rho]$. The standard choice is $w_{\lambda}(r_{ij}) =
\lambda/r_{ij}$, but other choices are possible. If the density
supplied to $F_{\lambda}[\rho]$ is held fixed at a
physical density generated by an appropriate (high-level, systematically improvable) \textit{ab-initio}
quantum-chemical methodology and if the value of the interaction-strength
parameter is varied from $0$ to $1$, then the adiabatic
connection~\cite{Harris:1974p1782,LANGRETH:1975p1425,GUNNARSSON:1976p1781,Gunnarsson:1977,LANGRETH:1977p1780}
between the Kohn--Sham ($\lambda=0$) and physical ($\lambda=1$)
systems can be studied numerically. Such studies for atomic and
molecular species~\cite{COLONNA_JCP110_2828,TEALE_JCP130_104111,TEALE_JCP132_164115,TEALE_JCP133_164112,STROMSHEIM_JCP135_194109}
can provide useful insight into the failings of standard
density-functional approximations and provide data for the construction and
evaluation of new forms, based on the modeling of the adiabatic
connection~\cite{BECKE_JCP98_1372,SEIDL_PRA59_51,SEIDL_PRA62_012502,PGG_JCTC5_743,COHEN_JCP127_034101,TEALE_JCP132_164115}.

Having established a convex formulation of CDFT in Sec.~\ref{sec:restoreconc}, it is possible to calculate the adiabatic
connection in CDFT in a manner similar to that of standard DFT. Details of the adiabatic connection for CDFT have been presented
previously by Liu~\cite{LIU_PRA54_1328}. The ground-state energy functionals at interaction strength $\lambda$ are given by
\begin{align}
  \label{eqELAMBDAv}
  E_{\lambda}[v,\mathbf{A}] & = \inf_{\Gamma} \trace{\Gamma H_{\lambda}[v,\mathbf{A}]}, \\
  \label{eqELAMBDAu}
  \bar{E}_{\lambda}[u,\mathbf{A}] & = E_{\lambda}[u-\tfrac{1}{2} A^2,\mathbf{A}],
\end{align}
where we have introduced the Hamiltonian
\begin{equation}
  H_{\lambda}[v,\mathbf{A}] = \frac{1}{2} \sum_k \pi_k^2 + \sum_k v(\mathbf{r}_k) + W_\lambda.
\label{HAMlambda}
\end{equation}
From the exact functional $\bar{E}_{\lambda}[u,\mathbf{A}]$ or an
accurate approximation to it, the adiabatic connection may be studied in terms the corresponding Vignale--Rasolt functional
\begin{equation}
  F_{\mathrm{VR},\lambda}[\rho,\mathbf{j}_{\text{p}}] = \sup_{u,\mathbf{A}} \left[ \bar{E}_{\lambda}[u,\mathbf{A}] - (\rho | u) - (\mathbf{j}_{\text{p}} | \mathbf{A})  \right],
  \label{ACF}
\end{equation}
which needs to be evaluated for a fixed density $(\rho,\mathbf{j}_{\text{p}})$
and different values of $\lambda$ in the interval $0\leq \lambda
\leq 1$. Typically, the density is the ground-state
density for some external potential
$(v_{\text{ext}},\mathbf{A}_{\text{ext}})$ at $\lambda=1$.
The optimization in Eq.~\eqref{ACF} is trivial for
$\lambda = 1$ since the optimal potential is then
$(v_{\mathrm{ext}}+\tfrac{1}{2}A_{\text{ext}}^2,
\mathbf{A}_{\mathrm{ext}})$; for $0 \leq \lambda < 1$,
the optimization is non-trivial. In particular, for $\lambda=0$, 
the optimal potential is the Kohn--Sham potential
$(u_{\text{s}}, \mathbf{A}_{\text{s}})$ given by
\begin{align}
u_{\text{s}} &=  v_{\text{ext}}   + v_{\text{J}}  + v_{\mathrm{xc}}  + \tfrac{1}{2} A_\text s^2 = v_{\text{s}}  + \tfrac{1}{2} A_\text s^2, \label{SCAPOT} \\
\mathbf{A}_{\text{s}} &= \mathbf{A}_{\text{ext}} + \mathbf{A}_{\text{xc}}, \label{VECPOT}
\end{align}
where the classical Coulomb or Hartree potential $v_{\text{J}}$ and the exchange--correlation potential $v_{\text{xc}}$ are
the functional derivatives of the corresponding energy
components as in standard Kohn--Sham DFT. The exchange--correlation
contribution to the vector potential is defined as
\begin{equation}
\mathbf{A}_{\mathrm{xc}} = \frac{\delta E_{\mathrm{xc}}[\rho,\mathbf{j}_{\text{p}}]}{\delta \mathbf{j}_{\text{p}}},
\end{equation}
where differentiability in relevant directions is assumed. These scalar and vector
potentials then enter the CDFT Kohn--Sham
equations~\cite{VIGNALE_PRL59_2360}, which may be re-written in terms
of $(u_{\text{s}}, \mathbf{A}_{\text{s}})$ as
\begin{equation}
\left[ \tfrac{1}{2} p^2 + \tfrac{1}{2} \{\mathbf{p}, \mathbf{A}_{\text{s}} \} +  u_{\text{s}}\right] \varphi_p = \varepsilon_p \varphi_p.
\end{equation}
If the spin dependent $\mathbf{B}(\mathbf{r}_k) \cdot
\mathbf{S}$ term is included in the Hamiltonian of Eq.~\eqref{HAM} with
the modified interactions of Eq.~\eqref{WLAM}, then similar arguments
apply. Two-component spinors rather than one-particle
orbitals then occur in the Kohn--Sham equations, allowing for a treatment of
non-collinear magnetism.

Even in the absence of external magnetic fields, violations of non-interacting $v$-representability---that is,
the existence of a ground-state density of the fully interacting Hamiltonian
$H_1[v,\mathbf{A}]$ that cannot be reproduced by
Slater-determinantal ground-states of the non-interacting Hamiltonian
$H_0[v_\text s,\mathbf{A}_\text s]$---have been shown to be common in two-electron
systems~\cite{TAUT_PRA80_022517}. In general, an extended Kohn--Sham
formalism, allowing for an ensemble description and fractional occupation numbers, is therefore
required in CDFT as well as in standard DFT.

To facilitate the optimization of
$F_{\mathrm{VR},\lambda}[\rho,\mathbf{j}_{\text{p}}]$ at a general interaction
strength $\lambda$, we restrict our attention to classes
of potentials that can be parameterized in a simple way---for example,
as linear combinations of basis functions. The most direct way to
benefit from the concavity of $\bar{E}[u,\mathbf{A}]$
is to parameterize $u$ rather than $v$ in the affine form
\begin{align}
  u(\mathbf{r}) & = u_{\text{ref}}(\mathbf{r}) + \sum_t b_t f_t(\mathbf{r}), \\
  \mathbf{A}(\mathbf{r}) & = \mathbf{A}_{\text{ref}}(\mathbf{r}) + \sum_t \mathbf{c}_t g_t(\mathbf{r}).
\end{align}
The use of $u$ rather than $v$ eliminates the $A^2$ term and the associated
quadratic dependence on $\mathbf{c}_t$, thereby simplifying the equations obtained upon substitution
in Eq.~\eqref{ACF}. Importantly, it also ensures that all stationary
points are true global maxima. To perform optimizations similar to those in
Refs.~\cite{WU_JCP118_2498,TEALE_JCP130_104111,TEALE_JCP132_164115,TEALE_JCP133_164112,STROMSHEIM_JCP135_194109},
all that remains is to be able to calculate the ground-state energy
$\bar{E}_\lambda[u,\mathbf{A}]$ with sufficient accuracy and to choose
appropriate basis functions $\{ f_t \}$ and $\{g_t\}$. Derivatives with respect to the expansion coefficients
$b_t$ and $\mathbf{c}_t$ may then be used in a quasi-Newton procedure
analogous to that in Ref.~\cite{WU_JCP118_2498}.
The choice of basis functions and reference potentials raises the issue
of a balanced descriptions of $u$ and $\mathbf{A}$, as well as the
asymptotic limits and reduction of gauge freedom inherent in the finite basis set.
Although important for practical implementation, these issues are beyond the scope of this article.

\subsection{A note on spin densities}

The formulation of CDFT with $(u,\mathbf{A})$ as the basic potential
and with the intrinsic and extrinsic energies expressed as mutual
Legendre--Fenchel transforms makes the introduction of spin straightforward.
The addition of the spin-Zeeman operator to the Hamiltonian introduces
an energy term containing the spin density $\mathbf{m}(\mathbf{r})$
paired with the magnetic field, $(\mathbf{m}|\mathbf{B}) = (\mathbf{m}|\nabla\times\mathbf{A})$.
A partial integration transfers the curl operator to the spin density
and gives a surface term if the integration domain is
finite. Neglect of the surface term leads to the energy term
\begin{equation}
  (\mathbf{m}|\nabla\times\mathbf{A}) = (\boldsymbol{\nabla} \times \mathbf{m} | \mathbf{A})
\end{equation}
and a theory with $(\rho,\mathbf{j}_{\text{m}})$ as the basic density. Here,
\begin{equation}
  \mathbf{j}_{\text{m}} = \mathbf{j}_{\text{p}} + \nabla\times\mathbf{m},
\end{equation}
is the sum of the paramagnetic current and the \emph{spin current}
$\nabla\times\mathbf{m}$. All results above remain valid with
$\mathbf{j}_{\text{p}}$ replaced by $\mathbf{j}_\text m$ and with
suitable modifications of the Hamiltonian and definitions of
$\mathcal{V}_N$ and $\mathcal{B}_N^{\text{p}}$.

The use of $\mathbf{j}_{\text{m}}$ as a basic density has been discussed by
Capelle and Gross~\cite{CAPELLE_PRL78_1872} as a way to translate
functionals between spin-density functional theory (SDFT) and CDFT. The
alternative formulation of current-spin DFT (CSDFT) in terms of the charge
density $\rho$, the spin density $\mathbf{m}$, and the paramagnetic current
density $\mathbf{j}_{\text{p}}$ as separate basic variables is a less
attractive formal theory since two independent potentials such as $(u,\mathbf{A})$
cannot be conjugate to three independent densities
$(\rho,\mathbf{m},\mathbf{j}_{\text{p}})$. On the other hand, construction of
practical approximate exchange--correlation functionals may be
substantially more difficult with the basic variables $\rho$ and $\mathbf{j}_\text m$.

\section{\label{secPHYSCDFT}The physical current as a basic variable}

Mathematically, it is not surprising that a theory formulated in terms
of magnetic vector potentials and wave functions---both gauge-dependent
objects---makes use of gauge-dependent basic variables. Indeed, with
gauge-dependent notions being so deeply entrenched in the theory, it
is not trivial to construct a useful reformulation that features only
gauge-invariant basic variables. On the other hand, from a physical point of
view, it is somewhat unappealing that the paramagnetic current, rather
than the physical current, arises as a basic density. Therefore, some authors
have attempted the construction of an alternative CDFT, with the physical current
as a basic variable. In particular, Pan and Sahni have gone far in
arguing that the paramagnetic current density, in some sense, cannot
correctly be regarded as a basic CDFT
variable~\cite{PAN_IJQC110_2833,PAN_JPCS73_630,SAHNI_PRA85_052502,PAN_PRA86_042502}.

\subsection{Do physical densities determine potentials up to a gauge?}

An important question that arises in CDFT is to what extent an HK
theorem is possible when the physical densities $(\rho,\mathbf{j})$
are chosen as the basic densities---that is, whether
$(v_1,\mathbf{A}_1)\notgaugeeq(v_2,\mathbf{A}_2)$ always implies that
$(\rho_1,\mathbf{j}_1)\neq(\rho_2,\mathbf{j}_2)$. 

Two simple observations lend plausibility to this claim.
First, for one-electron systems, $(v,\mathbf{A})$ can be found
explicitly given $(\rho,\mathbf{j})$. Writing the one-electron wave function as
\begin{equation}
  \psi(\mathbf{r}) = R(\mathbf{r}) \text e^{\text i S(\mathbf{r})},
\end{equation}
where $R$ and $S$ are real valued, we obtain the density $\rho = R^2$
and physical current density $\mathbf{j} = \rho(\boldsymbol{\nabla} S
+ \mathbf{A})$. Because of the identity
$\boldsymbol{\nabla}\times\boldsymbol{\nabla} S = \mathbf{0}$, the
physical densities $\rho$ and $\mathbf{j}$ determine the external
magnetic field,
\begin{equation}
\mathbf{B}(\mathbf{r}) = \boldsymbol{\nabla}\times\mathbf{A}(\mathbf{r}) = \boldsymbol{\nabla}\times \frac{\mathbf{j(\mathbf{r})}}{\rho(\mathbf{r})} .
\end{equation}
The scalar potential $v$ may be determined
up to a constant from the eigenvalue equation $(\tfrac{1}{2} \pi^2 +
v - E) \psi = 0$ and the observation that
$\boldsymbol{\pi} = -\text i \boldsymbol{\nabla} + \mathbf{A} = -\text i \boldsymbol{\nabla} + \mathbf{j}/\rho -  \boldsymbol{\nabla} S$, yielding
\begin{equation}
 \left[\frac{1}{2} \left(-\mathrm{i} \boldsymbol{\nabla} + \frac{\mathbf{j}(\mathbf{r})}{\rho(\mathbf{r})}
 \right)^2 + v(\mathbf{r}) - E \right] \rho^{1/2}(\mathbf{r}) = 0,
\end{equation}
from which $v(\mathbf{r}) - E$ is uniquely determined.
Hence, for a one-electron system, $\mathbf{B}$ and $v$ are both determined by $\rho$ and $\mathbf{j}$.
For an $N$-electron system, we obtain more generally
\begin{equation}
  \boldsymbol{\nabla}\times\frac{\mathbf{j(\mathbf{r})}}{\rho(\mathbf{r})} = \boldsymbol{\nu}(\mathbf{r}) + \mathbf{B}(\mathbf{r}),
\end{equation}
where $\boldsymbol{\nu} =
\boldsymbol{\nabla}\times\rho^{-1}\mathbf{j}_{\text{p}}$ is the
vorticity introduced by Vignale and
Rasolt~\cite{VIGNALE_PRL59_2360}. Vorticity is a gauge-invariant
quantity, but it is not clear whether it can be uniquely reconstructed
from $(\rho,\mathbf{j})$, without a priori knowledge of the external
magnetic field $\mathbf{B}$.

Following Pan and Sahni~\cite{PAN_IJQC110_2833}, the second
observation is that, while there is no HK theorem for CDFT in terms of
\emph{paramagnetic} densities, the counterexamples to the existence of an HK theorem for paramagnetic densities
(such as the harmonic-oscillator system in Sec.~\ref{sec:electronic})
do not preclude an HK theorem for physical densities. 

To see this, note that these counterexamples all exhibit 
different potentials $(v_1,\mathbf{A}_1) \notgaugeeq
(v_2,\mathbf{A}_2)$ with the same ground state $\psi_1 = \psi_2$ and
therefore the same paramagnetic density $(\rho_1,\mathbf{j}_{\text{p}1}) =
(\rho_2,\mathbf{j}_{\text{p}2})$. However, according to the original
HK theorem, this situation is impossible when
$\mathbf{A}_1 = \mathbf{A}_2$ since this would imply $v_2 \neq v_1+c$ and therefore 
$\rho_1\neq \rho_2$. We must therefore assume that $\mathbf{A}_1\neq \mathbf{A}_2$ in the counterexamples. 
However, from the
assumption $\mathbf{j}_{\text{p}1}=\mathbf{j}_{\text{p}2}$, it then follows that
$\mathbf{j}_1 - \mathbf{j}_2 = \rho_1 (\mathbf{A}_1 -
\mathbf{A}_2) \neq \mathbf{0}$.
In short, if $(v_1,\mathbf{A}_1) \notgaugeeq (v_2,\mathbf{A}_2)$
share the same ground-state wave function, then the \emph{physical
densities} are not the same.
It remains to explore whether it is possible to have different ground
states $\psi_1\neq \psi_2$ but the same physical densities
$(\rho_1,\mathbf{j}_1)=(\rho_2,\mathbf{j}_2)$; if possible, then no HK
theorem exists for the physical current densities.

General arguments for an HK theorem for physical
current densities have been put forth by Pan and
Sahni~\cite{PAN_IJQC110_2833,PAN_JPCS73_630} and by
Diener~\cite{DIENER_JP3_9417}. However, as discussed below, neither of
these arguments amounts to a rigorous proof. To our knowledge, the
existence of an HK theorem for physical current
densities therefore remains open.

\subsection{Pan and Sahni's argument}

The standard HK theorem of DFT states that $v\neq v'+c$ implies
$\rho'\neq \rho'$. The proof has two parts: first, it is shown that scalar
potentials that differ by more than a constant must have different
wave functions $\psi$ and $\psi'$; second, this result is combined with the
Rayleigh--Ritz variation principle to show that the assumptions
$\rho=\rho'$ and $v\neq v'+ c$ lead to a contradiction. It follows that $v\neq
v'+c$ implies $\rho\neq\rho'$.

For CDFT with physical densities, a proof along the same lines has
been attempted by Pan and Sahni~\cite{PAN_IJQC110_2833}. The first
part of their argument establishes that different potentials
$(v,\mathbf{A}) \notgaugeeq (v',\mathbf{A}')$ cannot yield the same
physical density $(\rho_0,\mathbf{j}_0)=(\rho'_0,\mathbf{j}'_0)$ if
the ground-state wave functions are the same. That is, without loss of
generality, it may be assumed in an HK-type argument that the wave
functions are different. The second part of their argument seeks to
establish, by the Rayleigh--Ritz variation principle, that
gauge-inequivalent potentials cannot have both different ground-state
wave functions and the same physical ground-state density.  To this
end, two potentials $(v,\mathbf{A}), (v',\mathbf{A}') \in
\mathcal{V}_N$, $(v,\mathbf{A}) \neq (v',\mathbf{A}')$, are considered
and it is argued that a contradiction arises from the following
assumptions:
\begin{itemize}
\item[(a)] The potentials have the same
physical density $(\rho_0,\mathbf{j}_0)$.
\item[(b)] The potentials
differ by more than a gauge transformation, $(v',\mathbf{A}')
\notgaugeeq \ (v,\mathbf{A})$.
\end{itemize}
No contradiction can result from (a)
alone, so the assumption (b) is crucial for any reductio ad absurdum
argument to succeed.  For example, because of gauge freedom, (a) and
the negation of (b) are certainly not contradictory since this
corresponds to the perfectly consistent situation where two potentials
that differ by a gauge transformation give rise to the same physical
density.  A correct proof must therefore contain at least one step
that makes use of (b). However, while stated as an assumption, (b) is
in fact never used in Pan and Sahni's argument. The argument must
consequently be invalid.

In more detail, we here identify an erroneous step in the reasoning in
Ref.~\cite{PAN_IJQC110_2833}, in particular Eqs.~(35)--(40). With
$\psi$ and $\psi'$ denoting the ground states of the two potentials,
we obtain Eq.~(35) from Ref.~\cite{PAN_IJQC110_2833}:
\begin{equation}
  \label{eqPSINEQ}
  E = \bra{\psi} H[v,\mathbf{A}] \ket{\psi} < \bra{\psi'} H[v,\mathbf{A}] \ket{\psi'}.
\end{equation}
The identity $H[v,\mathbf{A}] = H[v',\mathbf{A}'] + (H[v,\mathbf{A}] - H[v',\mathbf{A}'])$
yields for the expectation value on the right-hand side 
\begin{widetext}
\begin{equation}
 \begin{split}
  \bra{\psi'} H[v,\mathbf{A}] \ket{\psi'} & = E' + \bra{\psi'} \tfrac{1}{2} \pi^2 + v - \tfrac{1}{2} \pi'^2 - v' \ket{\psi'}   \\
    & = E' + \bra{\psi'} \tfrac{1}{2} \{\boldsymbol{\pi},\mathbf{A}\} + v - \tfrac{1}{2} A^2 - \tfrac{1}{2} \{\boldsymbol{\pi}',\mathbf{A}'\} - v' + \tfrac{1}{2} A'^2 \ket{\psi'}.
   \label{HExp}
 \end{split}
\end{equation}
\end{widetext}
It is here important to distinguish between $\boldsymbol{\pi} = \mathbf{p}
+ \mathbf{A}$ and $\boldsymbol{\pi}' = \mathbf{p}+\mathbf{A}'$ since
the representation of the mechanical momentum operator is gauge
and vector-potential dependent.  To proceed, we explicitly
write out the corresponding physical current density operators, which
for the two potentials are given by
\begin{align}
  \op{\mathbf{j}}(\mathbf{r}) & = \frac{1}{2} \sum_k (\boldsymbol{\pi}_k \delta(\mathbf{r}_k-\mathbf{r}) + \delta(\mathbf{r}_k-\mathbf{r}) \boldsymbol{\pi}_k), \\
  \op{\mathbf{j}}'(\mathbf{r}) & = \frac{1}{2} \sum_k (\boldsymbol{\pi}'_k \delta(\mathbf{r}_k-\mathbf{r}) + \delta(\mathbf{r}_k-\mathbf{r}) \boldsymbol{\pi}'_k).
\end{align}
From these, we may calculate the physical current (assumed to be the same in the two cases) and its interaction with some vector potential $\mathbf{a}(\mathbf{r})$
as
\begin{align}
  \mathbf{j}_0(\mathbf{r}) &= \bra{\psi} \op{\mathbf{j}}(\mathbf{r}) \ket{\psi} = \bra{\psi'} \op{\mathbf{j}}'(\mathbf{r}) \ket{\psi'}  \\
  (\mathbf{j}_0 | \mathbf{a})
&= \bra{\psi}  \tfrac{1}{2} \{ \boldsymbol{\pi}, \mathbf{a} \} \ket{\psi}
= \bra{\psi'}  \tfrac{1}{2} \{ \boldsymbol{\pi}', \mathbf{a} \} \ket{\psi'} ,
\end{align}
where it is important to use primed or unprimed quantities consistently. To
ensure that we are using the correct current-density operator for
$\psi'$ in Eq.~\eqref{HExp}, we insert the identity
$\{\boldsymbol{\pi},\mathbf{A}\} = \{\boldsymbol{\pi}' +
\mathbf{A}-\mathbf{A}',\mathbf{A}\}$ yielding,
\begin{equation}
 \label{eqHAMAPRIME}
 \begin{split}
E < E' + (\rho_0|\Delta v) + (\mathbf{j}_0|\Delta \mathbf{A})  + \tfrac{1}{2} (\rho_0|\Delta A^2).
 \end{split}
\end{equation}
where we have introduced $\Delta v = v - v'$ and $\Delta \mathbf{A} = A - A'$ and also used the inequality in Eq.~\eqref{eqPSINEQ}.

Carrying out the above argument with primed and unprimed variables interchanged, we obtain the strict inequality
\begin{equation}
  \label{eqPSINEQ2}
  E + E' < E' + E + (\rho_0|\Delta A^2),
\end{equation}
where the last term does not vanish since $\Delta A^2$ is symmetric in the primed and unprimed variables. In agreement with the HK
theorem, a contradiction arises if (and only if) $\mathbf{A} = \mathbf{A}'$. When $\mathbf{A}(\mathbf{r})$ and $\mathbf{A}'(\mathbf{r})$ differ where
$\rho(\mathbf{r}) \neq 0$ (as a result of different physical fields or by a gauge transformation), no contradiction arises.

In the argument given in Ref.~\cite{PAN_IJQC110_2833}, the
authors incorrectly identify $\bra{\psi'} \op{\mathbf{j}}(\mathbf{r})
\ket{\psi'}$ and $\bra{\psi'} \op{\mathbf{j}}'(\mathbf{r})
\ket{\psi'}$, leading to their Eq.~(38), which differs from
Eq.~\eqref{eqHAMAPRIME} above by the replacement of $\tfrac{1}{2}
(\rho_0|\Delta A^2)$ with $\tfrac{1}{2}( \rho_0 | A^{\prime 2} - A^2
)$. Since their term is antisymmetric rather than symmetric in the
primed and unprimed variables, the authors obtain a contradiction $E +
E' < E + E'$ irrespective of $\mathbf{A}$ and $\mathbf{A}'$, leading
to the unjustified conclusion that ($\rho_0,\mathbf{j}_{0})$ determines
$(v,\mathbf{A})$.

\subsection{Diener's argument}

An earlier attempt to prove an HK-type theorem for physical currents
by Diener~\cite{DIENER_JP3_9417} invokes an intriguing strategy for
eliminating the term $(\rho_0|\Delta A^2)$ from
Eq.~\eqref{eqPSINEQ2}. The key idea is to replace the external vector
potential $\mathbf{A}$ by an effective vector potential
$\mathbf{a}_{\text{eff}}[\mathbf{j},\psi]$
defined to reproduce a given physical current density $\mathbf{j}$ from a given wave function $\psi$ so that
\begin{equation}
  \mathbf{j} = \mathbf{j}_{\text{p};\psi} + \rho_{\psi} \mathbf{a}_{\text{eff}}[\mathbf{j},\psi].
\end{equation}
Consider now the following rearrangement of the kinetic energy term
\begin{align}
  \pi^2 & = p^2 + \{\mathbf{p}, \mathbf{A}\} + A^2 \nonumber \\
        & = p^2 - a_{\text{eff}}^2 + \{\mathbf{p}+\mathbf{a}_{\text{eff}},\mathbf{A}\} + (\mathbf{a}_{\text{eff}}-\mathbf{A})^2.
\end{align}
By definition, $\frac{1}{2} \bra{\psi}
\{\mathbf{p}+\mathbf{a}_{\text{eff}}[\mathbf{j},\psi],\mathbf{A}\} \ket{\psi} =
(\mathbf{j}|\mathbf{A})$. For a prescribed physical current density
$\mathbf{j}_0$, the expectation value of an arbitrary wave function is
then
\begin{align}
  \bra{\psi} H[v,\mathbf{A}] \ket{\psi} & = \bra{\psi} \tfrac{1}{2} (p^2 - a_{\text{eff}}^2) + W \ket{\psi}
\nonumber \\
   &
+ (\mathbf{j}_0|\mathbf{A}) + (\rho|v) + \tfrac{1}{2}(\rho | (\mathbf{a}_{\text{eff}}-\mathbf{A})^2).
 \label{eqHWITHEFFA}
\end{align}
The total current evaluated using the effective
momentum operator $\mathbf{p}+\mathbf{a}_{\text{eff}}$ instead of the true
mechanical momentum operator $\mathbf{p}+\mathbf{A}$ always
evaluates to the prescribed current $\mathbf{j}_0$.
Diener then defines a universal density functional of the form
\begin{align}
  F_{\text{D}}[\rho,\mathbf{j}] & = \inf_{\psi\mapsto\rho} \bra{\psi} H_{\text{eff}}[\mathbf{j},\psi] \ket{\psi},
         \\
   H_{\text{eff}}[\mathbf{j},\psi] & = \tfrac{1}{2} p^2 - \tfrac{1}{2} a_{\text{eff}}[\mathbf{j},\psi]^2 + W. \label{DienerH}
\end{align}
By exploiting this functional, Diener derives the inequality
\begin{equation}
  \label{eqDIENERINEQ}
  \bra{\psi_0} H[v,\mathbf{A}] \ket{\psi_0} \leq \bra{\psi'_0} H[v,\mathbf{A}] \ket{\psi'_0} - \frac{1}{2} (\rho_0 | \Delta A^2),
\end{equation}
in lieu of the usual strict Rayleigh--Ritz inequality underlying the HK proof as given in~Eq.~\eqref{eqPSINEQ}.
When $\psi'_0$ is the ground state of potentials that differ by more
than a gauge, $(v',\mathbf{A}') \notgaugeeq (v,\mathbf{A})$, the above
non-strict inequality is (without further ado) taken to be strict in
Diener's presentation. If a strict inequality is accepted, 
a standard reductio ad absurdum proof is possible because
the last term of Eq.~\eqref{eqPSINEQ2} cancels to yield the
contradiction $E + E' < E' + E$.

We now consider two technical problems not addressed by Diener. The
first is that the expectation value being minimized in
$F_{\text{D}}$ is not bounded from below for densities that vanish at
some point. Although unusual, such ground-state densities can arise
for small molecules in strong magnetic fields.  Consider a wave
function $\Phi_0$ giving rise to a density $\rho_0$ that vanishes
at some point in space $\mathbf{O}$.  Let us also introduce spherical coordinates $(r,\theta,\phi)$ about
$\mathbf{O}$. Then, for the wave functions
\begin{equation}
  \Phi_m(\mathbf{x}_1,\ldots,\mathbf{x}_N) = \text e^{2\pi \text{i} m\sum_{k=1}^N \phi_k} \Phi_0(\mathbf{x}_1,\ldots,\mathbf{x}_N)
\end{equation}
with integer $m$, we see that $\Phi_m$ and $\Phi_0$ give rise to the same density $\rho_0$ but
to different paramagnetic current densities related by
\begin{equation}
  \mathbf{j}_{\text{p};m} = \frac{m \hat{\boldsymbol{\phi}}}{r \sin(\theta)} \rho_0 + \mathbf{j}_{\text{p};0},
\end{equation}
where $\hat {\boldsymbol \phi}$ is the unit vector in the direction specified by $\phi$.
As a result,
\begin{equation}
  \mathbf{a}_m = \mathbf{a}_{\text{eff}}[\mathbf{j}, \Phi_m] = \mathbf{a}_{\text{eff}}[\mathbf{j}, \Phi_0] - \frac{m \hat{\boldsymbol{\phi}}}{r \sin(\theta)} = \mathbf{a}_0 - \frac{m \hat{\boldsymbol{\phi}}}{r \sin(\theta)}.
\end{equation}
Calculating the expectation value of the effective Hamiltonian in
Eq.~\eqref{DienerH}, some terms arising from $\tfrac{1}{2} p^2$ cancel
terms arising from $\tfrac{1}{2} a_m^2$, leaving
\begin{align}
  \bra{\Phi_m} \tfrac{1}{2} (p^2 - a_m^2) + W \ket{\Phi_m} & =
   \bra{\Phi_0} \tfrac{1}{2} (p^2-a_0^2) + W \ket{\Phi_0} \nonumber \\
       &  \ \  + m \int \frac{\hat{\boldsymbol{\phi}}\cdot \mathbf{j}}{r \sin(\theta)} \, \mathrm{d}\mathbf{r}
.
\end{align}
For physical currents $\mathbf{j}$ with a nonzero last integral, for
example a circular current, the expectation value can be decreased
without bound, demonstrating that $F_{\text{D}}[\rho,\mathbf{j}]$ is
not well defined on the full domain $\mathcal{B}_N$.

The second technical problem in the derivation of the inequality in
Eq.~\eqref{eqDIENERINEQ} concerns the strategy of choosing a
prescribed $\mathbf{j}_0$ such that the effective vector potential
equals the external vector potential. This amounts to a
self-consistency condition [Eq.~(11) of Ref.~\onlinecite{DIENER_JP3_9417}],
\begin{equation}
  \mathbf{j}_0 = \mathbf{j}_{\text{p};\Phi_{\text{D}}} + \rho_0 \mathbf{a}_{\text{eff}}[\mathbf{j}_0,\Phi_{\text{D}}] = \mathbf{j}_{\text{p};\Psi[v,\mathbf{A}]} + \rho_0 \mathbf{A},
\end{equation}
where $\Phi_{\text{D}}$ minimizes
$\bra{\Phi_{\text{D}}} H_{\text{eff}}[\mathbf{j}_0,\Phi_{\text{D}}]
\ket{\Phi_{\text{D}}}$ and $\Psi[v,\mathbf{A}]$ is the ground state of the
external potentials giving rise to $(\rho_0,\mathbf{j}_0)$. Diener's
derivation of the inequality in Eq.~\eqref{eqDIENERINEQ} hinges on the
ability to reproduce the paramagnetic current density from a ground
state $\Psi[v,\mathbf{A}]$ of $H[v,\mathbf{A}]$ by a minimizing wave function
in $F_{\text{D}}[\rho_0,\mathbf{j}_0]$. However, it is unclear whether this is always
possible.

\subsection{Constrained search with physical currents}


In lieu of a rigorous proof for an HK-type theorem for physical currents,
it is possible to proceed by conjecturing such a result and exploring
its consequences. Additionally, an important point is that a mapping
from ground state densities to potentials may not be required for a
formulation of CDFT, provided that the theory can be constructed by
some other means such as a constrained search or Legendre--Fenchel
transformation formalism. We shall here explore such issues in the framework
introduced by Pan and
Sahni~\cite{PAN_IJQC110_2833,PAN_JPCS73_630,SAHNI_PRA85_052502,PAN_PRA86_042502} in
more detail.

A complication due to the choice of the physical current as
a basic variable is that constraints of the type ``$\psi \to \rho,
\mathbf{j}$'' require explicit reference to a vector potential,
because the physical current is not determined by the wave function
alone. Care must therefore be exercised when developing a constrained-search
formalism for physical currents.

For each magnetic field under consideration, we fix a gauge. Hence, we choose a mapping
\begin{equation}
\mathbf{B}(\mathbf{r}) \mapsto
\mathbf{a}[\mathbf{B}](\mathbf{r})
\end{equation}
from magnetic fields to magnetic vector potentials, and also fix the
constant shift of scalar potentials in some way.
From the conjecture that a ground state density $(\rho,\mathbf{j})$
uniquely determines a gauge class of potentials
$\overline{(v,\mathbf{A})}$, we may now write mappings
\begin{equation}
  (\rho,\mathbf{j}) \leftrightarrow (\nabla v,\mathbf{B}) \leftrightarrow (v,\mathbf{a}[\mathbf{B}])
\end{equation}
Hence, we may write the scalar potential, the external magnetic field and its vector potential
as functionals $v = w[\rho,\mathbf{j}]$, $\mathbf{B} =
\mathbf{b}[\rho,\mathbf{j}]$, and $\mathbf{A} =
\mathbf{a}[\mathbf{b}[\rho,\mathbf{j}]]$ of the physical densities; these are
representatives of the equivalence class $\overline{(v,\mathbf{A})}$.

Given potentials $(v,\mathbf{A})$, we may determine the corresponding
ground state $\psi_0$ and physical ground-state densities
$(\rho,\mathbf{j}) \in \mathcal{B}_N$, and express the energy as
\begin{equation}
 \begin{split}
  E[v, \mathbf{A}] = \bra{\psi_0} \tfrac{1}{2} p^2 + W \ket{\psi_0} + (\rho|v-\tfrac{1}{2} A^2) + (\mathbf{j}|\mathbf{A}).
 \end{split}
\end{equation}
Using the mapping from densities to potentials, a universal density
functional could in principle internally reconstruct the ground state
potentials, and obtain the intrinsic energy via
\begin{equation}
 \begin{split}
  F_{\text{PS}}[\rho,\mathbf{j}] & = E\left[w[\rho,\mathbf{j}],\mathbf{a}[\mathbf{b}[\rho,\mathbf{j}]]\right]
- (\rho\,|\,w[\rho,\mathbf{j}]-\tfrac{1}{2} a[\mathbf{b}[\rho,\mathbf{j}]]^2) \\
     & \ \  - (\mathbf{j}\,|\,\mathbf{a}[\mathbf{b}[\rho,\mathbf{j}]]).
 \end{split}
\end{equation}
Furthermore, we may define a non-universal constrained-search
functional that depends explicitly on the external vector potential,
\begin{equation}
 \begin{split}
 F^{\text{cs1}}_{\text{PS}}[\rho,\mathbf{j},\mathbf{A}] & =
\inf_{\Gamma\mapsto(\rho,\mathbf{j}-\rho\mathbf{A})} \trace{\Gamma (\tfrac{1}{2} p^2 + W)}  \\
    &  = F_{\text{VR}}[\rho,\mathbf{j}-\rho\mathbf{A}],
 \end{split}
\end{equation}
where we note that $\mathbf{j} - \rho \mathbf{A} = \mathbf{j}_{\text{p}}$.
In principle, given the conjectured one-to-one mapping between
potentials and physical densities, a functional of physical densities
may internally reconstruct the corresponding potentials. This observation is
relied on in the definition of $F_{\text{PS}}$ and it is
tempting to consider also a universal constrained-search functional that exploits this idea,
\begin{equation}
 \begin{split}
 F^{\text{cs2}}_{\text{PS}}[\rho,\mathbf{j}] & =
\inf_{\Gamma\mapsto(\rho,\mathbf{j}-\rho\mathbf{a}[\mathbf{b}[\rho,\mathbf{j}]])} \trace{\Gamma (\tfrac{1}{2} p^2 + W)}  \\
    &  = F_{\text{VR}}[\rho,\mathbf{j}-\rho\mathbf{a}[\mathbf{b}[\rho,\mathbf{j}]]].
 \end{split}
\end{equation}

Consider now the reformulation of the minimization in Eq.~\eqref{eqCDFTVFUN} as a nested
minimization over physical densities and wave functions.
\begin{equation}
 \begin{split}
  \label{eqPSCONSEARCH}
  E[v,\mathbf{A}] & = \inf_{\rho',\mathbf{j}'} \Big[ \overbrace{\inf_{\Gamma\mapsto\rho',\mathbf{j}'-\rho'\mathbf{A}} \trace{\Gamma (\tfrac{1}{2} p^2 + W)}}^{F^{\text{cs1}}_{\text{PS}}[\rho',\mathbf{j}',\mathbf{A}] } \\
   & \ \ + (\rho'|v-\tfrac{1}{2} A^2) + (\mathbf{j}'|\mathbf{A})  \Big]
   \end{split}
\end{equation}
A constrained-search-like expression is thus possible, but the
functional that arises from the nested minimization is the
non-universal functional $F^{\text{cs1}}_{\text{PS}}$. A minimization
over the universal functional $F^{\text{cs2}}_{\text{PS}}$ is here not
equivalent,
\begin{equation}
 \begin{split}
  \label{eqPSCONSEARCH2}
  E[v,\mathbf{A}] & \neq \inf_{\rho',\mathbf{j}'}  \Big[ \overbrace{ \inf_{\Gamma\mapsto\rho',\mathbf{j}'-\rho'\mathbf{a}[\mathbf{b}[\rho',\mathbf{j}']]} \trace{\Gamma (\tfrac{1}{2} p^2 + W)}}^{F^{\text{cs2}}_{\text{PS}}[\rho',\mathbf{j}']} \\
     & \ \ + (\rho'|v-\tfrac{1}{2} A^2) + (\mathbf{j}'|\mathbf{A})  \Big].
   \end{split}
\end{equation}
To establish the non-equality, consider the introduction of a new
scalar potential $u_{-} = v - \frac{1}{2} A^2$. Written in terms of
$u_{-}$, the right-hand side of Eq.~\eqref{eqPSCONSEARCH2} has the
form of a Legendre--Fenchel transformation and consequently a
functional of $(u_{-},\mathbf{A})$ that is concave by
construction. However, writing the harmonic oscillator energy of Sec.~\ref{sec:restoreconc} in terms of this scalar potential,
we see
that the resulting energy functional is not concave in
$(u_{-},\mathbf{A})$ (it is in fact only concave in scalar
potentials $u_{\lambda} = v + \lambda A^2$ for $\lambda \ge \frac{1}{2}$).
 Hence this non-concave energy functional
cannot equal the concave functional defined in terms of
$F^{\text{cs2}}_{\text{PS}}$. In previous work (see, e.g.,
Eqs.~(24)--(28) of Ref.~\cite{PAN_JPCS73_630}), these two
non-equivalent functionals were conflated.

The constrained-search approach to CDFT with the physical current as a
basic variable is substantially complicated by the fact that a wave
function does not determine the physical current. It does not appear
possible to construct a functional that is both universal and admits a
straightforward variation principle. Moreover, unlike the situation
when $\mathbf{j}_{\text{p}}$ is a basic variable, a simple redefinition of the
scalar potential does not simultaneously yield a concave energy
functional and linear pairing between potentials and physical
densities. An avenue for further study could be to consider the properties of a
universal intrinsic energy functional of the physical
densities. Such a functional cannot be convex in the pair
$(\rho,\mathbf{j})$ as a whole, but could possibly be convex in $\rho$
(for a fixed $\mathbf{j}$) and concave in $\mathbf{j}$ (for a fixed
$\rho$), which would enable at least partial Legendre--Fenchel
transformations.

\section{Conclusions}\label{sec:conc}

In this work we have compared formulations of CDFT based on different
choices of basic variables. While the usual formulation in terms of
$(\rho, \mathbf{j}_{\text{p}})$ and $(v,\mathbf{A})$ does not lead to a
one-to-one mapping between potentials and densities, it allows for the
construction of a useful density functional theory via a constrained
search approach. A drawback with this choice of basic variables is
that $E[v,\mathbf{A}]$ is not concave, making a full Legendre--Fenchel
transform treatment infeasible. However, as shown here, concavity can
be restored by considering the conjugate variables $(\rho,
\mathbf{j}_{\text{p}})$ and $(u,\mathbf{A})$, where $u=v+\frac{1}{2}A^2$. This
allows the application of convex analysis in analogy with Lieb's
formulation of standard DFT~\cite{LIEB_IJQC24_243}. Such a formulation
is particularly natural in the context of the study of adiabatic
connections for CDFT functional construction, which can be done in a
manner similar to that undertaken for standard
DFT~\cite{COLONNA_JCP110_2828,TEALE_JCP130_104111,TEALE_JCP132_164115,TEALE_JCP133_164112,STROMSHEIM_JCP135_194109}. The
information garnered from such analysis would be suitable for
comparison with Kohn--Sham implementations of CDFT based on
functionals of $(\rho,\mathbf{j}_{\text{p}})$.

Alternative CDFT formulations based on the physical densities
$(\rho,\mathbf{j})$ have also been critically examined. Pan and
Sahni's~\cite{PAN_IJQC110_2833,PAN_JPCS73_630,SAHNI_PRA85_052502,PAN_PRA86_042502}
recent attempt to formulate such a theory is found to be
unsatisfactory. Both their attempt to prove the existence of a
one-to-one mapping between potentials and densities and their
constrained-search formulation were found to be flawed. An earlier
attempt to formulate a CDFT in terms of the physical current by
Diener~\cite{DIENER_JP3_9417} has also been examined, and technical
difficulties with this approach have been highlighted. Despite the
appealing physical motivation behind this choice of basic variables,
we thus find that a formal justification for such a framework is
currently lacking. Furthermore, while it remains open whether or not
an analogue of the Hohenberg--Kohn theorem holds for the physical
current, other aspects of standard DFT such as the variation
principle, the constrained-search formalism, and formulations in terms
of Legendre--Fenchel transformations do not straightforwardly carry
over to this type of CDFT. We conclude that the most common
formulation in terms of $(\rho, \mathbf{j}_{\text{p}})$ is presently the most
convenient and viable formulation of CDFT.

\section*{Appendix}

Let $X$ be a normed vector space and $X^\ast$ its dual---that is, the the linear space of all continuous linear functionals on $X$.
A function $f: X \to \mathbb R \cup \{+\infty\}$ is said to be convex if it satisfies the relation
\begin{equation}
f(\lambda x_1 + (1-\lambda) x_1) \leq \lambda f(x_1) + (1-\lambda) f(x_2)
\end{equation}
for all $0 \leq \lambda \leq 1$. The effective domain, $\dom(f)$, is the set of $x \in X$ for which $f(x) < +\infty$.
A function $f$ is lower semi-continuous at $x_0$ if, for any $\epsilon > 0$, there exists $\delta >0$ such that
$f(x) \geq f(x_0) - \epsilon$ whenever $| x - x_0 | < \delta$; $f$ is lower semi-continuous
if it is lower semi-continuous at all $x \in X$.

A lower semi-continuous convex function $f: X \to \mathbb R \cup \{+\infty\}$
may be represented by its conjugate $f^\ast: X^\ast \to \mathbb R \cup \{+\infty\}$ in the manner
\begin{align}
f^\ast(y) &= \sup_{x\in X} [(x | y) - f(x)], \\
f(x) &= \sup_{y \in X^\ast} [(x | y) - f^\ast(y)],
\end{align}
where $f^\ast$ is also lower semi-continuous and convex. The dual function
$y_0 \in X^\ast$ is said to be a subgradient of $f$ at $x_0 \in X$ if it satisfies the inequality
\begin{equation}
f(x) \geq f(x_0) + (x - x_0 | y_0), \quad \forall x \in X. \label{subgradient}
\end{equation}
The set of all subgradients of $f$ at $x_0$ is called the subdifferential $\subdiff f(x_0)$ and is a (possibly empty) convex subset of $X^\ast$.
Subgradients and subdifferentials of $f^\ast$ are defined in an analogous manner.
The function $f$ and its conjugate satisfy Fenchel's inequality $f(x) + f^\ast(y) \geq ( x | y)$, which is sharpened into
the equality $f(x) + f^\ast(y) = ( x | y)$ whenever the equivalent reciprocal relations
\begin{equation}
y \in \subdiff f(x) \iff x \in \subdiff f^\ast(y)
\end{equation}
are satisfied.

A function $f: X \to \mathbb{R} \cup \{-\infty\}$ is said to be concave if $-f$ is convex. Its conjugate is defined in the same manner as for
convex functions but with $\sup$ replaced by $\inf$; likewise, subgradients and subdifferentials are defined as for a convex
function but with the inequality sign reversed in Eq.~\eqref{subgradient}.

\begin{acknowledgments}
  This work was supported by the Norwegian Research Council through
  the CoE Centre for Theoretical and Computational Chemistry (CTCC)
  Grant No.\ 179568/V30 and the Grant No.\ 171185/V30 and through the
  European Research Council under the European Union Seventh Framework
  Program through the Advanced Grant ABACUS, ERC Grant Agreement No.\
  267683. A. M. T. is also grateful for support from the Royal Society University Research Fellowship scheme.
\end{acknowledgments}


%

\end{document}